\documentclass[preprint,notoc]{JHEP3}
\usepackage{epsfig,amsmath}

\def\to{\rightarrow}

\def\bi{\begin{itemize}}
 \def\ei{\end{itemize}}

\def\c1p{C1^\prime}
\def\ta{\tilde a}
\def\tG{\tilde G}

\def\ta{\tilde a}

\def\tst{\tilde t}
\def\ttau{\tilde \tau}

\def\tw{\widetilde W}
\def\tz{\widetilde Z}
\def\alt{\stackrel{<}{\sim}}
\def\agt{\stackrel{>}{\sim}}
\def\be{\begin{equation}}  
\def\ee{\end{equation}}  
\def\bea{\begin{eqnarray}}  
\def\eea{\end{eqnarray}}  

\newcommand\njp[3]{{\it New\ J.\ Phys.\ }{\bf #1} (#2) #3}

\newcommand\sjp[3]{{\it Sov.\ J.\ Nucl.\ }{\bf #1} (#2) #3}

\title{Fine-tuning favors mixed axion/axino cold dark matter\\
over neutralinos in the minimal supergravity model}
\author{Howard Baer and  Andrew D. Box\\
$^a$Dept.\ of Physics and Astronomy, University of Oklahoma, Norman, OK 73019, USA\\
E-mail: \email{baer@nhn.ou.edu}, \email{box@nhn.ou.edu}}

\preprint{\vbox{}}

\abstract{
Over almost all of minimal supergravity (mSUGRA or CMSSM) model parameter space, 
there is a large overabundance of neutralino cold dark matter (CDM).
We find that the allowed regions of mSUGRA parameter space which match the
measured abundance of CDM in the universe are highly fine-tuned.
If instead we invoke the Peccei-Quinn-Weinberg-Wilczek solution 
to the strong $CP$ problem, 
then the SUSY CDM may consist of an axion/axino admixture with an axino mass of order
the MeV scale, and where mixed axion/axino or mainly axion CDM seems preferred. 
In this case, 
fine-tuning of the relic density is typically much lower, showing that axion/axino CDM 
($a\tilde{a}$CDM) 
is to be preferred in the paradigm model for SUSY phenomenology.
For mSUGRA with $a\tilde{a}$CDM, quite different regions of parameter
space are now DM-favored as compared to the case of neutralino DM. 
Thus, rather different SUSY signatures are expected at the LHC 
in the case of mSUGRA with $a\tilde{a}$CDM, as compared to mSUGRA with neutralino CDM. 
}

\begin{document}

\section{Introduction}
\label{sec:intro}

A wide array of astrophysical data point to us living in a universe comprised of
$4\%$ baryons, $\sim 25\%$ cold dark matter (CDM) and $\sim 70\%$ dark energy.
In fact, the cosmic abundance of CDM has been recently measured
to high precision by the WMAP collaboration\cite{wmap5}, which finds
\be
\Omega_{CDM}h^2=0.110\pm 0.006 ,
\ee
where $\Omega=\rho/\rho_c$ is the dark matter density relative to the 
closure density, and $h$ is the scaled Hubble constant.
No particle present in the Standard Model (SM) of particle physics 
has the correct properties to constitue the CDM, so some form of new physics 
is needed. It is compelling, however, that candidate CDM particles do emerge naturally 
from two theories which provide solutions to longstanding problems in particle physics. 

The first problem-- known as the gauge hierarchy problem-- arises due
to quadratic divergences in the scalar sector of the SM. 
These divergences
lead to scalar masses blowing up to the highest scale in the theory
({\it e.g.} in grand unified theories (GUTS), 
the GUT scale $M_{GUT}\simeq 2\times 10^{16}$ GeV), unless
an enormous fine-tuning of parameters is invoked. 
One solution to the gauge hierarchy problem occurs by introducing supersymmetry (SUSY) into the
theory. The inclusion of softly broken SUSY leads to
a cancellation of quadratic divergences between fermion and boson loops, 
so that only log divergences remain. 
The log divergence is soft enough that
vastly different scales remain stable within a single effective theory.
In SUSY theories, the lightest neutralino emerges as an excellent
WIMP CDM candidate. 
Gravity-mediated SUSY breaking models (supergravity, or SUGRA) contain gravitinos with weak-scale masses. 
SUGRA models experience tension due to
possible overproduction of gravitinos in the early universe, leading to an overabundance
of CDM.
In addition, gravitinos usually decay during or after Big Bang nucleosynthesis (BBN),
and their energetic decay products may disrupt
the successful calculations of light element
abundances, which otherwise maintain good agreement with observation.
This tension in SUGRA models is known as the {\it gravitino problem}.

The second problem is the strong $CP$ problem\cite{kcreview}. An elegant solution 
to the strong $CP$ problem was proposed by Peccei and Quinn (PQ) many years ago\cite{pq}.
The PQ solution automatically predicts the existence of a new particle (WW)\cite{ww}: 
the axion $a$. 
While the original PQWW axion was soon ruled out, models of a nearly
``invisible axion'' were developed in which the PQ symmetry breaking scale
was moved up to energies of order $f_a\sim 10^{9}-10^{12}$ GeV\cite{ksvz,dfsz}.
The axion also turns out to be an excellent candidate 
particle for CDM in the universe\cite{absik}.

Of course, it is highly desirable to simultaneously account for
{\it both} the strong $CP$ problem and the gauge hierarchy problem.
In this case, it is useful to invoke supersymmetric models which
include the PQWW solution to the strong $CP$ problem\cite{nillesraby}. In a 
SUSY context, the axion field is just one element of an 
{\it axion supermultiplet}. The axion supermultiplet contains 
a complex scalar field, whose real part is the $R$-parity even saxion 
field $s(x)$, and whose imaginary part is the axion field $a(x)$.
The supermultiplet also contains an $R$-parity odd spin-$\frac{1}{2}$ 
Majorana field, the axino $\ta$\cite{steffen_rev}.
The saxion, while being an $R$-parity even field, nonethless 
receives a SUSY breaking mass likely of order the weak scale. 
The axion mass is constrained by
cosmology and astrophysics to lie in a favored range 
$10^{-2}$ eV$\agt m_a\agt 10^{-5}$ eV. 
The axino mass is very model dependent\cite{axmass,rtw,cl,ckkr}, depending
heavily on the exact form of the superpotential and the mechanism for SUSY breaking. 
In supergravity models, it may be of
order the gravitino mass $m_{3/2}\sim$ TeV, or as low as $m_{3/2}^2/f_a\sim $keV.
Conditions for realizing these extremes are addressed in \cite{cl}. 
Here, we will try to avoid explicit model-dependence, and adopt 
$m_{\ta}$ as lying within the general range of keV-GeV, 
as in numerous previous works\cite{rtw,ckkr,fstw,cmssm,axdm}.
An axino in this mass range would likely serve as the lightest
SUSY particle (LSP), and is also a good candidate particle for
cold dark matter\cite{rtw,ckkr}.

In a previous paper\cite{axdm}, we investigated supersymmetric models wherein the PQ
solution to the strong $CP$ problem is assumed. For definiteness,
we restricted the analysis to examining the paradigm minimal
supergravity (mSUGRA or CMSSM) model\cite{msugra}. 
We were guided in our analysis by considering the possibility of including 
a viable mechanism for baryogenesis in the early universe. 
In order to do so, we needed
to allow for re-heat temperatures after the inflationary epoch to 
reach values $T_R\agt 10^6$ GeV. We found that in order to sustain
such high re-heat temperatures, as well as generating predominantly 
{\it cold} dark matter, we were pushed into mSUGRA parameter
space regions that are very different from those allowed by
the case of thermally produced neutralino dark matter. In addition, we
found that very high values of the PQ breaking scale $f_a/N$ of order
$10^{11}-10^{12}$ GeV were needed, leading to the mSUGRA model with
{\it mainly axion cold dark matter}, but also with a small
admixture of thermally produced axinos, and an even smaller
component of warm axino dark matter arising from neutralino decays.
The favored axino mass value is of order 100 keV.
We note here recent work on models with dominant axion CDM explore the 
possibility that axions form a cosmic Bose-Einstein condensate, which can 
allow for the solution of several problems associated with large scale
structure and the cosmic background radiation\cite{pierre}.

In this paper, we will examine the mSUGRA model under the assumption 1. of
neutralino CDM and 2. that 
mixed axion/axino DM ($a\tilde{a}$DM) saturates the WMAP measured 
abundance\footnote{The possibility of mixed $a\tilde{a}$CDM was suggested
in the context of Yukawa-unified SUSY in Ref. \cite{mix}.}.
To compare the two DM scenarios, we will evaluate a measure of
fine-tuning in the relic abundance
\be
\Delta_{a_i} \equiv\frac{\partial\log\Omega_{DM}h^2}{\partial\log a_i}
\ee
with respect to variations in fundamental parameters $a_i$ of the model.
Such a measure of relic abundance fine-tuning was previously calculated in Ref.~\cite{eo} 
in the context of just neutralino dark matter.
Here, we will expand upon this and also consider fine-tuning of the relic
density in the case of mixed $a\tilde{a}$DM.
Our main conclusion is that the relic abundance of DM is {\it much less fine-tuned
in the case of mixed $a\tilde{a}$CDM, as compared to neutralino CDM}.
Thus, we find that mixed $a\tilde{a}$CDM is {\it theoretically preferable to 
neutralino CDM}, at least in the case of the mSUGRA model, and probably also
in many cases of SUGRA models with non-universal soft SUSY breaking terms. 

We will restrict our work to cases where the lightest neutralino
$\tz_1$ is either the LSP or the next-to-lightest SUSY particle (NLSP) with an
axino LSP; the case with a stau NLSP and an axino LSP has recently been examined in Ref. \cite{fstw}.
Related previous work on axino DM in mSUGRA can be found in Ref. \cite{cmssm}.

The remainder of this paper is organized as follows.
In Sec. \ref{sec:inoDM}, we calculate the neutralino relic abundance fine-tuning 
parameter $\Delta_{\tz_1}$ in the mSUGRA model due to variation in parameters
$m_0$ and $m_{1/2}$. 
We find, in good agreement with Ref. \cite{eo},
that the WMAP allowed regions are all finely-tuned for low values of $\tan\beta$. 
For much higher $\tan\beta\sim 50$, the fine-tuning is much less with respect to $m_0$
and $m_{1/2}$, but nevertheless high with respect to $\tan\beta$.
In Sec. \ref{sec:inoprob}, we review the gravitino problem, leptogenesis and the 
cosmological production of axion and axino dark matter.
In Sec. \ref{sec:axDM}, we calculate the fine-tuning parameter
$\Delta_{a\ta}$ for mixed $a\tilde{a}$CDM under the assumption of a very light
axino with $m_{\ta}\sim 0.1-1$ MeV. The fine-tuning is always quite low,
for both cases of mixed axino/axion CDM and mainly axion CDM. In the case of mainly axino
CDM, we find the scenario less well-motivated since for high values of 
$T_R\agt 10^6$ GeV, the value of $m_{\ta}\ll 0.1$ MeV, making the axino 
mainly {\it warm} DM instead of cold DM.
In Sec. \ref{sec:conclude}, we present a summary and conclusions.

\section{Fine-tuning in mSUGRA with neutralino cold dark matter}
\label{sec:inoDM}

\subsection{Overview}

We adopt the mSUGRA model\cite{msugra} as a template model for examining the
issue of fine-tuning in cases of neutralino CDM vs. $a\tilde{a}$CDM.
The mSUGRA parameter space is given by
\be
m_0,\ m_{1/2},\ A_0,\ \tan\beta ,\ sign (\mu ) ,
\ee
where $m_0$ is the unified soft SUSY breaking (SSB) scalar mass 
at the GUT scale, 
$m_{1/2}$ is the unified gaugino mass at $M_{GUT}$, $A_0$ is
the unified trilinear SSB term at $M_{GUT}$ and $\tan\beta\equiv
v_u/v_d$ is the ratio of Higgs field vevs at the weak scale.
The GUT scale gauge and Yukawa couplings, and the SSB terms are
evolved using renormalization group equations (RGEs) from
$M_{GUT}$ to $m_{weak}$, at which point electroweak symmetry is
broken radiatively, owing to the large top quark Yukawa coupling. 
At $m_{weak}$, the various sparticle and Higgs boson mass matrices are 
diagonalized to find the physical sparticle and Higgs boson masses.
The magnitude, but not the sign, of the superpotential $\mu$ parameter 
is determined by the EWSB minimization conditions.

We adopt the Isasugra subprogram of Isajet to 
generate sparticle mass spectra\cite{isajet}. 
Isasugra performs an iterative solution of the MSSM two-loop RGEs, and includes
an RG-improved one-loop effective potential evaluation at an optimized scale,
which accounts for leading two-loop effects\cite{haber}. Complete
one-loop mass corrections for all sparticles and Higgs boson masses
are included\cite{pbmz}. For the neutralino relic density, we use the IsaReD
subprogram of Isajet\cite{isared}.

Our measure of fine-tuning in the neutralino relic density, $\Delta_{\tz_1}$, 
is calculated by constructing a grid of points in $m_0-m_{1/2}$ space. 
At each point, the change in $\Omega_{\tz_1}h^2$ corresponding to a change in either 
$m_0$ or $m_{1/2}$ is calculated for both a positive and negative parameter change, using
\begin{equation}
\Delta_{a_i}=\frac{a_i}{\Omega_{\tilde{Z}_1}h^2}\frac{\partial \Omega_{\tilde{Z}_1}h^2}{\partial a_i}=
\frac{a_i}{\Omega_{\tilde{Z}_1}h^2}
\frac{\left[\Omega_{\tilde{Z}_1}h^2(a_i\pm\Delta a_i)-\Omega_{\tilde{Z}_1}h^2(a_i)\right]}{\Delta a_i}\;,
\end{equation}
where $a_i=m_0$ or $m_{1/2}$. For each $a_i$, the largest $\Delta_{a_i}$ is selected from the results for both 
the positive and negative change. 
To construct the overall total $\Delta_{\tz_1}$, the individual values are added in quadrature:
\begin{equation}
\Delta_{\tz_1} =\sqrt{\Delta_{m_0}^2+\Delta_{m_{1/2}}^2}\;.
\end{equation}
We may also consider fine-tuning due to variation in $A_0$ and $\tan\beta$. 
Variation in $A_0$ yields tiny variations in $\Omega_{\tz_1}h^2$ unless 
one moves close to the stop co-annihilation region (see Fig. \ref{fig:a0} in Sec. \ref{ssec:a0tanb}). 
Variation in $\tan\beta$ gives a slight effect on the relic density unless $\tan\beta$ becomes
very large. In this case, $m_A$ decreases\cite{ltanb} to the extent that $m_A\sim 2m_{\tz_1}$, and
neutralino annihilation rates are greatly increased due to the $A$-resonance.
Then, variation in $\tan\beta$ mainly shifts the {\it location} of 
the $A$-resonance in the $m_0\ vs.\ m_{1/2}$ plane. 
Moving on and off the resonance is already accounted for by varying 
$m_0$ and $m_{1/2}$. 
Nevertheless, in Sec. \ref{ssec:a0tanb} we present results due to including $A_0$ and $\tan\beta$ 
in the fine-tuning calculation.\footnote{
We note that Ref. \cite{eo} consider
fine-tuning versus variation in $m_b$ and $m_t$. We consider these as fixed
SM parameters, much as $M_Z$ is fixed.}

\subsection{Results from variation of $m_0$ and $m_{1/2}$}
\label{ssec:results}

Our first results are shown in Fig. \ref{fig:ino10}, where we show in frame {\it a}).
contours of $\Omega_{\tz_1}h^2$ in the $m_0\ vs.\ m_{1/2}$ mSUGRA plane for
$A_0=0$, $\tan\beta =10$ and $\mu >0$. We also take $m_t=172.6$ GeV. The well-known red regions 
are excluded  either due to a stau LSP (left-side) or lack of appropriate EWSB (lower and right side).
The gray-shaded region is excluded by LEP2 chargino searches ($m_{\tw_1}>103.5$ GeV), and
the green shaded region denotes allowable points with $\Omega_{\tz_1}h^2\le 0.11$. The region below 
the orange dashed contour is excluded by LEP2 Higgs searches, which require $m_h>114.4$ GeV; here, 
we actually require $m_h>111$ GeV to reflect a roughly 3 GeV error on the RGE-improved one-loop
effective potential calculation of $m_h$. 
\FIGURE[t]{
\includegraphics[width=14cm]{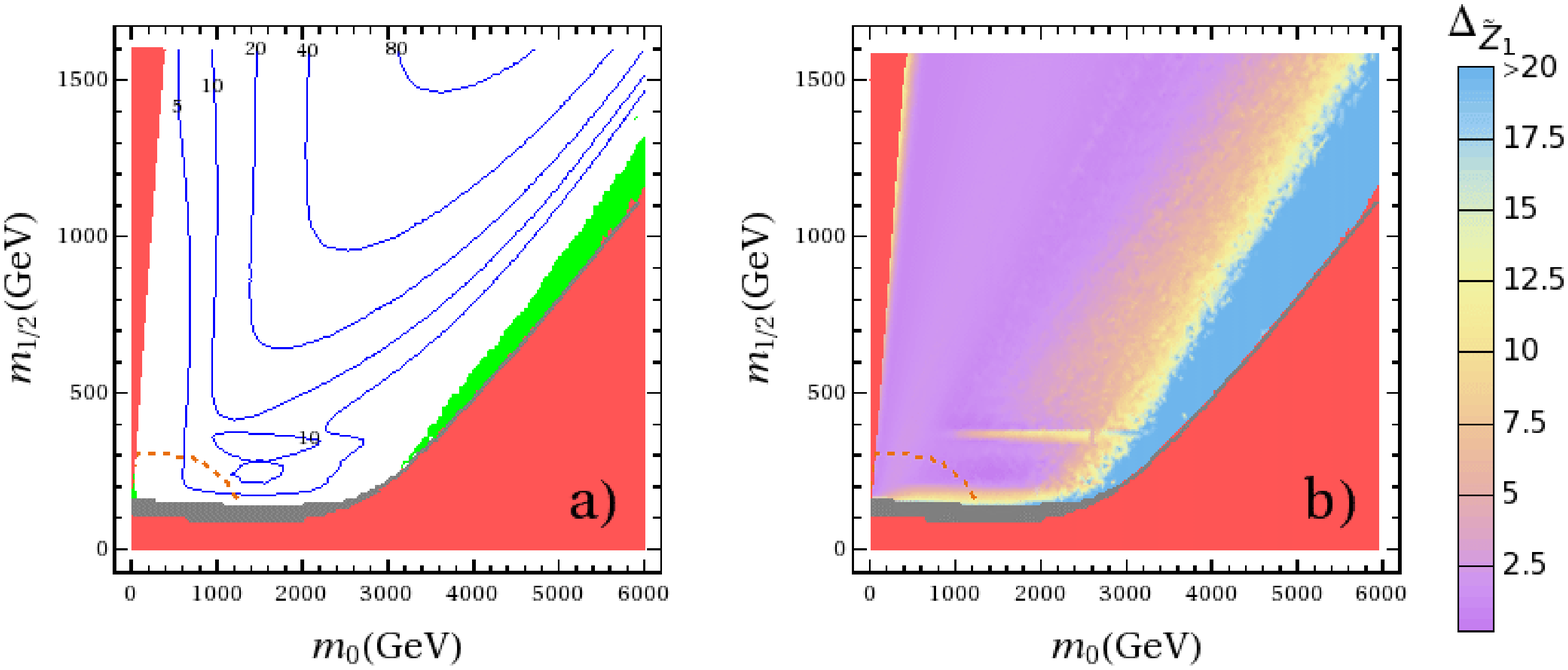}
\caption{In the $m_0\ vs.\ m_{1/2}$ plane of the mSUGRA model for
$A_0=0$, $\tan\beta =10$ and $\mu >0$, we plot
{\it a}). contours of $\Omega_{\tz_1}h^2$ and {\it b}). regions of 
fine-tuning parameter $\Delta_{\tz_1}$.
}\label{fig:ino10}}

The well-known (green-shaded) hyperbolic branch/focus point (HB/FP) region\cite{hb_fp} stands out on the 
right side, where $\mu$ becomes small and the $\tz_1$ becomes a mixed bino-higgsino state.
On the left edge, the very slight stau co-annihilation region\cite{stau} 
is barely visible.
We also plot contours of $\Omega_{\tz_1}h^2$ ranging from 5 to 80. In most of the 
mSUGRA parameter space, the relic abundance is 1-3 orders of magnitude higher than the 
WMAP measured value. The valley in $\Omega_{\tz_1}h^2$ around $m_{1/2}\sim 400$ GeV
is due to the turn-on of the $\tz_1\tz_1\to t\bar{t}$ annihilation mode.

In Fig. \ref{fig:3dn}, we show the neutralino relic density as a 3-d plot in the
$m_0\ vs.\ m_{1/2}$ plane, to gain extra perspective. The level of fine-tuning corresponds
to the slope of the surface. We see that in most of parameter space, the slope is relatively small,
{\it i.e.} the plateau is nearly flat. However, in this region, the relic density is far too high.
In the regions where $\Omega_{\tz_1}h^2\sim 0.1$, then the slope is extremely steep,
corresponding to large fine-tuning: a small variation in fundamental parameters leads to a large
change in relic density.
\FIGURE[t]{
\includegraphics[width=14cm]{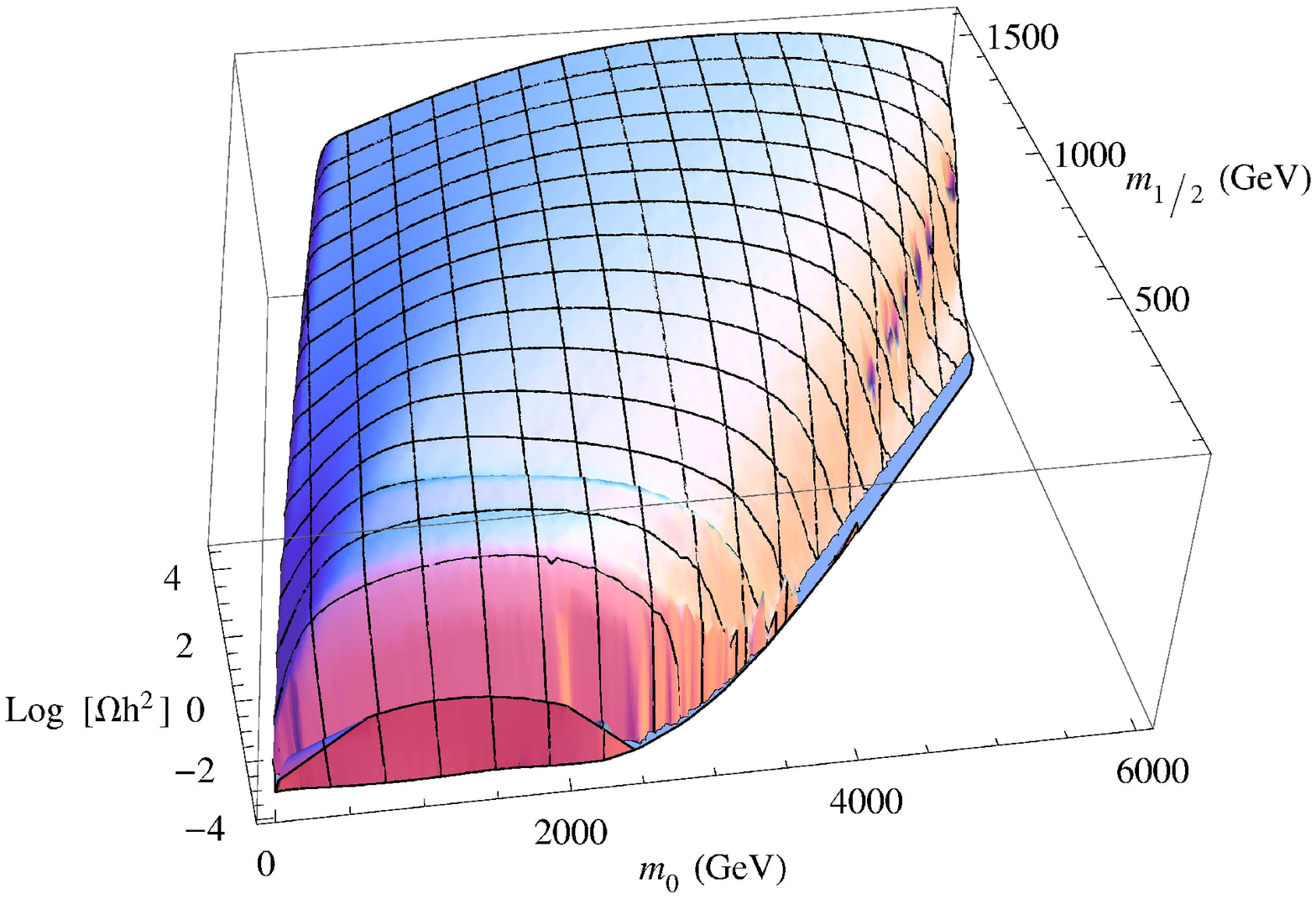}
\caption{A 3-d plot of neutralino relic density in the $m_0\ vs.\ m_{1/2}$ 
plane of the mSUGRA model for
$A_0=0$, $\tan\beta =10$ and $\mu >0$.
}\label{fig:3dn}}

In Fig. \ref{fig:ino10} {\it b})., we show regions of fine-tuning parameter $\Delta_{\tz_1}$. 
A value of $\Delta_{\tz_1}\sim 0$ corresponds to no fine-tuning (a flat slope in
$\Omega_{\tz_1}h^2$ versus variation in all parameters), while higher values of
$\Delta_{\tz_1}$ give increased fine-tuning in the relic density. We see immediately from the figure
that the vast majority of parameter space, where $\Omega_{\tz_1}h^2$ is much too large, 
is also not very fine-tuned. However, the HB/FP region, where $\mu\to 0$, has a very
high fine-tuning, with $\Delta_{\tz_1}$ ranging from 20-100! There are also regions of
substantial fine-tuning adjacent to the LEP2 chargino mass excluded region,
due to rapid changes in $\Omega_{\tz_1}h^2$ as one approaches the $\tz_1\tz_1\to h$ 
annihilation resonance\cite{hfunnel}, 
and also some fine-tuning at the turn on of $\tz_1\tz_1\to t\bar{t}$.
Finally, we see a very narrow region of fine-tuning extending along the stau co-annihilation
region.

To get a better grasp, we plot in Fig. \ref{fig:slice10} a slice out of
parameter space at $m_{1/2}=250$ and 500 GeV, showing in {\it a}).
$\Omega_{\tz_1}h^2$ and in {\it b}). $\Delta_{\tz_1}$ versus $m_0$.
We see the slope in $\Omega_{\tz_1}h^2$ is very steep in the HB/FP region, 
leading to $\Delta_{\tz_1}\sim 30$ (50) for lower (higher) $m_{1/2}$ values.
In contrast, in the stau co-annihilation region, where $\Omega_{\tz_1}h^2\sim 0.11$,
the value of $\Delta_{\tz_1}\sim 3$ (12) for low (high) $m_{1/2}$. The lower
$m_{1/2}$ value has only moderate fine-tuning since it is getting close to the 
``bulk'' annihilation region\cite{bulk}, where $\tz_1\tz_1$ annihilation is enhanced via light
$t$-channel slepton exhange diagrams.
\FIGURE[t]{
\includegraphics[width=14cm]{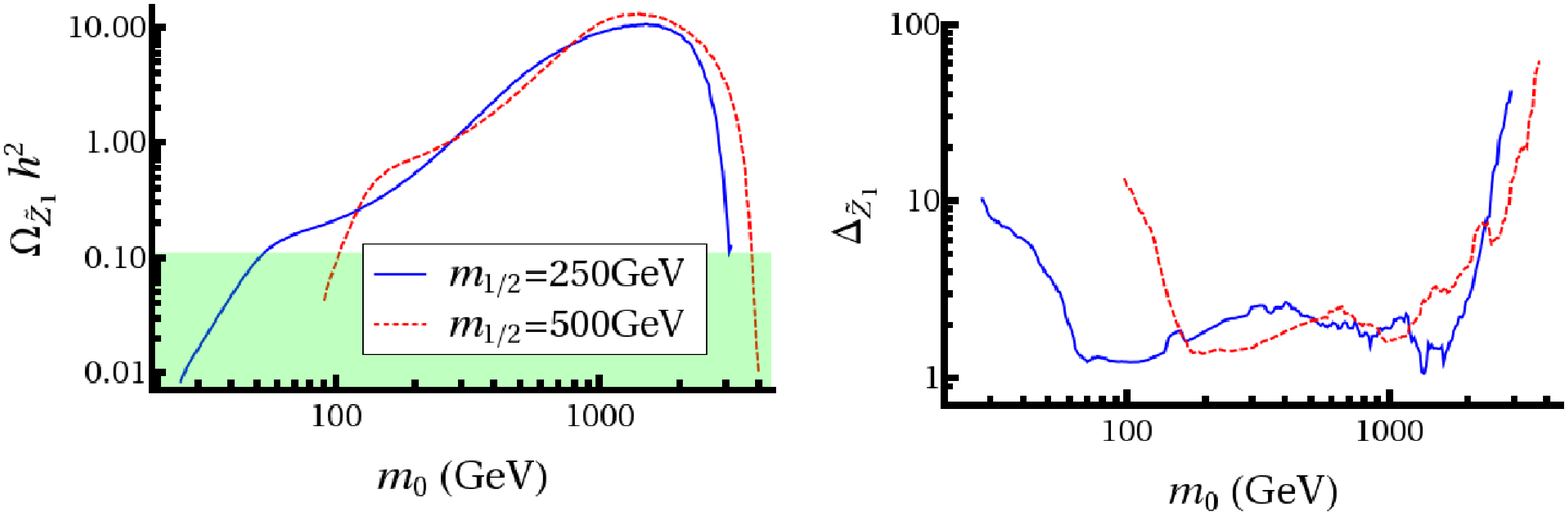}
\caption{A plot of {\it a}). $\Omega_{\tz_1} h^2$ and {\it b}).
$\Delta_{\tz_1}$ versus $m_0$ for fixed values of
$m_{1/2}=250$ GeV (blue) and $m_{1/2}=500$ GeV (red), in mSUGRA
with $A_0=0$, $\tan\beta =10$ and $\mu >0$.
}\label{fig:slice10}}

To gain a better perspective on the stau co-annihilation region, in Fig. \ref{fig:zoom10}
we show a blown-up portrait of the low $m_0$ region of parameter space. The ``turn-around''
in the green-shaded WMAP allowed region in frame {\it a}). is due to the impact of the bulk
annihilation region. Most of this area lies below the $m_h=111$ GeV contour, and thus gives rise 
to Higgs bosons that are too light. In frame {\it b}). is a blow-up of the fine-tuning parameter
$\Delta_{\tz_1}$. We see that the major portion of the stau co-annihilation region is
fine-tuned, with the {\it possible exception} of the region lying just below the
LEP2 $m_h$ bound, where mixed bulk/co-annihilation occurs.
\FIGURE[t]{
\includegraphics[width=14cm]{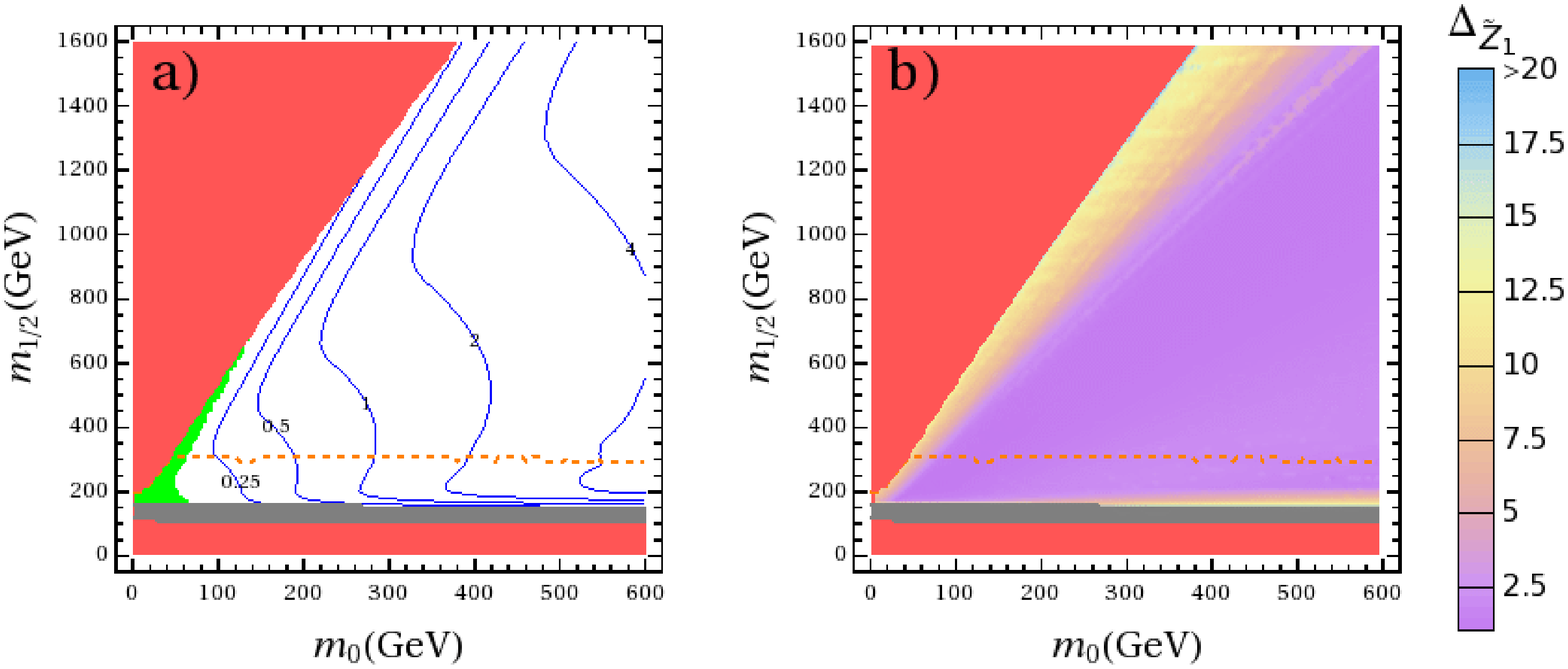}
\caption{A plot of {\it a}). contours of $\Omega_{\tz_1} h^2$ and {\it b}).
$\Delta_{\tz_1}$ in mSUGRA
with $A_0=0$, $\tan\beta =10$ and $\mu >0$.
This plot zooms in on the stau co-annihilation region.
}\label{fig:zoom10}}

In Fig. \ref{fig:ino30}, we show contours of relic density and $\Delta_{\tz_1}$
for $\tan\beta =30$. At higher values of $\tan\beta$, the $b$ and $\tau$ Yukawa couplings
increase in magnitude, and enhance neutralino annihilation into $b\bar{b}$ and $\tau\bar{\tau}$
final states. Overall, we see a similar picture to that shown in Fig. \ref{fig:ino10}
in that the HB/FP region has extreme fine-tuning of the relic density, while the stau
co-annihilation region is also fine-tuned, but somewhat less so. The bulk annihilation region
is again excluded by the LEP2 $m_h$ bound.
\FIGURE[t]{
\includegraphics[width=14cm]{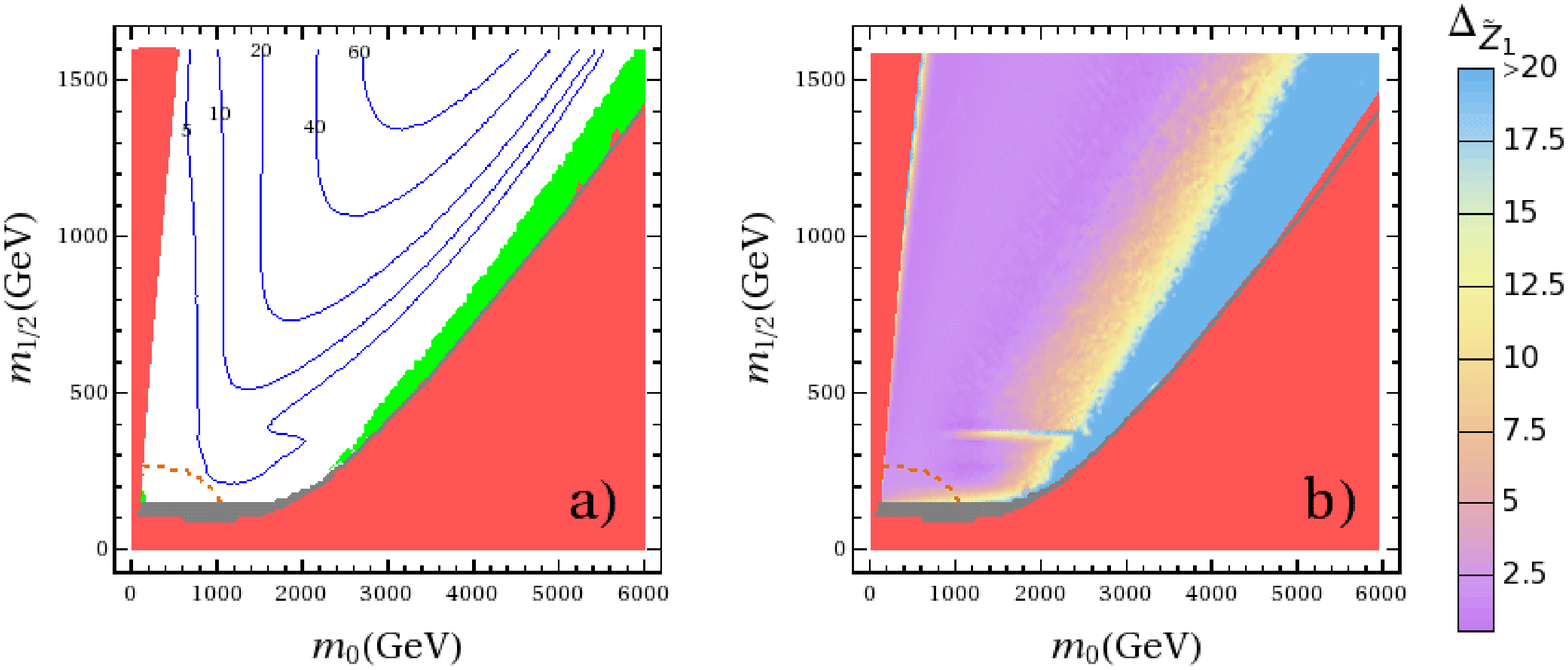}
\caption{In the $m_0\ vs.\ m_{1/2}$ plane of the mSUGRA model for
$A_0=0$, $\tan\beta =30$ and $\mu >0$, we plot
{\it a}). contours of $\Omega_{\tz_1}h^2$ and {\it b}). regions of 
fine-tuning parameter $\Delta_{\tz_1}$.
}\label{fig:ino30}}

We plot in Fig. \ref{fig:ino53} the mSUGRA plane for $\tan\beta =53$. In this case, a large
new green-shaded region is opening up along the low $m_0$ edge of parameter space. This is due to
three effects occuring at large $\tan\beta$. 1. The $\ttau_1$ mass decreases with $\tan\beta$, 
leading to increased annihilation into $\tau\bar{\tau}$ final states; 
this increases the area of the bulk annihilation region. 2. The tau and $b$ Yukawa couplings 
$f_\tau$ and $f_b$ increase, thus enhancing annihilation into $\tau\bar{\tau}$ and
$b\bar{b}$ final states. 
3. The value of $m_A$ is decreasing while the width $\Gamma _A$ is increasing 
(due to increasing Yukawa couplings that enter the $A$ decay modes), so that
$\tz_1\tz_1\to A^{(*)}\to b\bar{b},\ \tau\bar{\tau}$ increases: {\it i.e.} we are entering
the $A$ resonance annihilation region\cite{Afunnel}, 
which enhances the neutralino annihilation cross section
in the early universe, thus lowering the relic density. In the case of $\tan\beta =53$, we see
that the HB/FP region is still highly fine-tuned. However, broad portions of the low $m_0$ 
mSUGRA parameter space around $m_{1/2}\sim 300-600$ have $\Delta_{\tz_1}\alt 3$ due to
an overlap of bulk annihilation through staus, stau co-annihilation and $A$-resonance annihilation.
Another low fine-tuning and relic density consistent region occurs at $m_{1/2}\sim 1200$ GeV,
where one sits atop the $A$-resonance.
\FIGURE[t]{
\includegraphics[width=14cm]{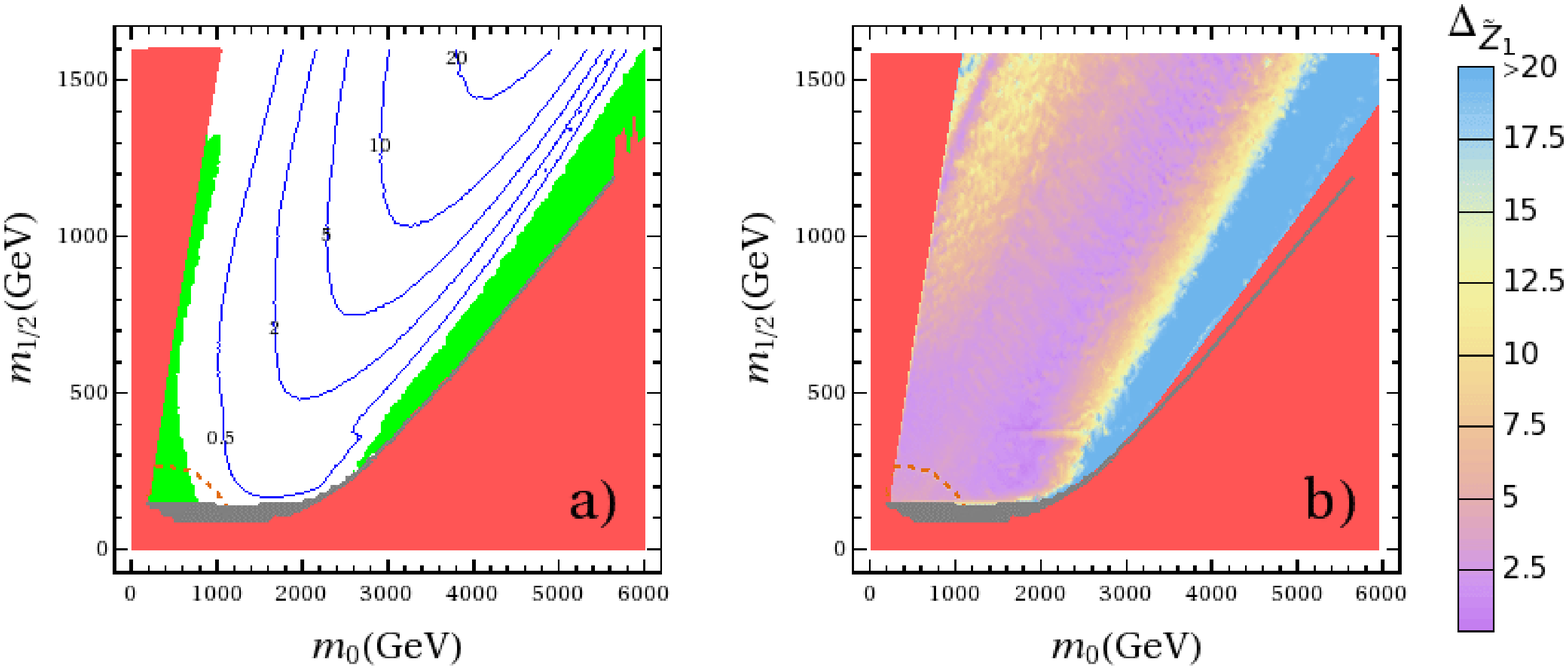}
\caption{In the $m_0\ vs.\ m_{1/2}$ plane of the mSUGRA model for
$A_0=0$, $\tan\beta =53$ and $\mu >0$, we plot
{\it a}). contours of $\Omega_{\tz_1}h^2$ and {\it b}). regions of 
fine-tuning parameter $\Delta_{\tz_1}$.
}\label{fig:ino53}}

In Fig. \ref{fig:ino55}, we show the mSUGRA plane for $\tan\beta =55$. Here, the
$A$-resonance annihilation region is fully displayed, and the $A$ width is even larger.
While much of the HB/FP region is still very fine-tuned, the regions of
annihilation though the broad $A$ resonance yield relatively low fine-tuning,
especially if one sits right on the resonance, or sits in the
resonance/bulk/co-annihilation overlap region at low
$m_0$ and low $m_{1/2}$.
\FIGURE[t]{
\includegraphics[width=14cm]{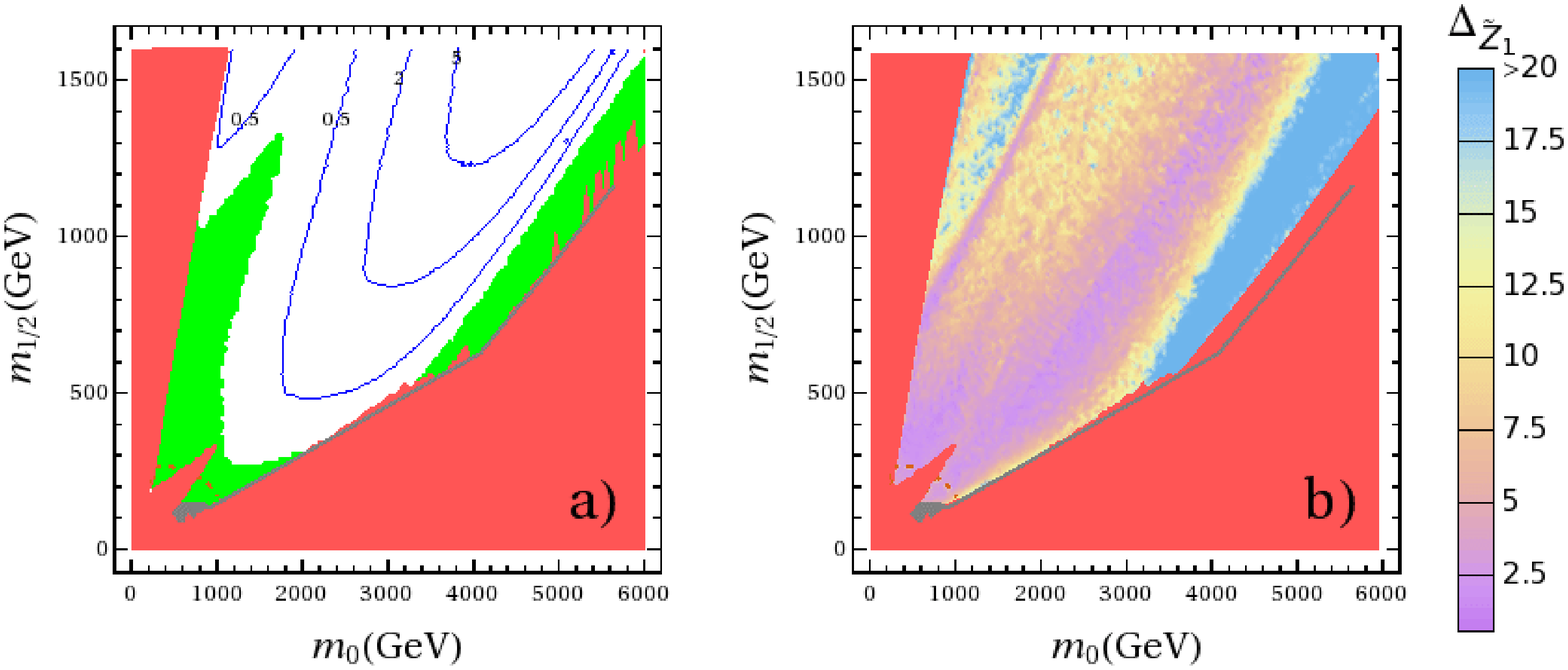}
\caption{In the $m_0\ vs.\ m_{1/2}$ plane of the mSUGRA model for
$A_0=0$, $\tan\beta =55$ and $\mu >0$, we plot
{\it a}). contours of $\Omega_{\tz_1}h^2$ and {\it b}). regions of 
fine-tuning parameter $\Delta_{\tz_1}$.
}\label{fig:ino55}}

\subsection{Results from variation in $A_0$ and $\tan\beta$}
\label{ssec:a0tanb}

As mentioned earlier, including $A_0$ into the measure of fine-tuning typically yields only a small
effect, unless one is near the top-squark co-annihilation region. This is because variation in $A_0$
mainly leads to different mixing in the third generation scalar system, and for most of 
mSUGRA parameter space, affects mainly the top squark mass eigenstates. To show this explicitly,
we plot in Fig. \ref{fig:a0}{\it a}). the value of $\Omega_{\tz_1}h^2$ and in frame {\it b}).
the value of $|\Delta_{A_0}|$ versus variation in $A_0$ for two cases: 
 1. $m_0=1.5$ TeV and $m_{1/2}=250$ GeV (blue curves) and 2. $m_0=2$ TeV and 
$m_{1/2}=750$ GeV (red dashed curve), for $\tan\beta =10$ and $\mu >0$.
In these two cases, the slope of $\Omega_{\tz_1}h^2$ is rather mild, leading to a 
contribution to $|\Delta_{A_0}|$ of typically 1 or less. The exception comes for the
$m_{1/2}=750$ GeV curve around $A_0\sim -4$ TeV, where indeed the value of $m_{\tst_1}$
is rapidly becoming lighter, and feeding into the relic density calculation.
\FIGURE[t]{
\includegraphics[width=14cm]{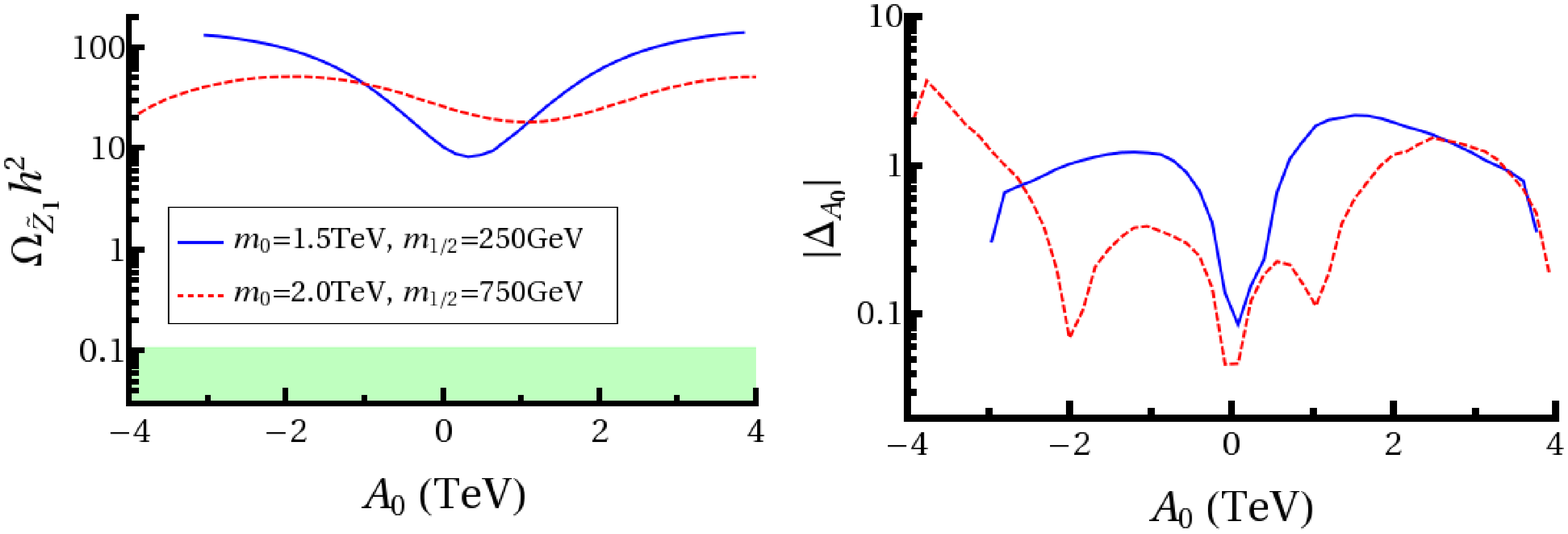}
\caption{Plot of {\it a}). $\Omega_{\tz_1}h^2$ and {\it b}). $|\Delta_{A_0}|$ 
versus $A_0$ for two slices out of mSUGRA parameter space: 
1. $m_0=1.5$ TeV and $m_{1/2}=250$ GeV (blue curves) and 2. $m_0=2$ TeV and 
$m_{1/2}=750$ GeV (red dashed curve), for $\tan\beta =10$ and $\mu >0$.
}\label{fig:a0}}

In Fig. \ref{fig:tanb}, we show {\it a}). the relic density and {\it b}). $|\Delta_{\tan\beta}|$
versus variation in $\tan\beta$ for 
 1. $m_0=1.5$ TeV and $m_{1/2}=250$ GeV (blue curves) and 2. $m_0=2$ TeV and 
$m_{1/2}=750$ GeV (red dashed curve), for $A_0=0$ and $\mu >0$. 
For most of the $\tan\beta$ values, the relic density varies only slowly with $\tan\beta$, 
leading to only small contributions to $\Delta$. When $\tan\beta$ becomes of order 50,
then $m_A$ is rapidly decreasing, and $\Gamma_A$ is rapidly increasing, leading to
a high rate of neutralino annihilation through the $A^0$ resonance. In this case, while
fine-tuning with respect to $m_0$ and $m_{1/2}$ is low, fine-tuning with respect to $\tan\beta$
is high.
\FIGURE[t]{
\includegraphics[width=14cm]{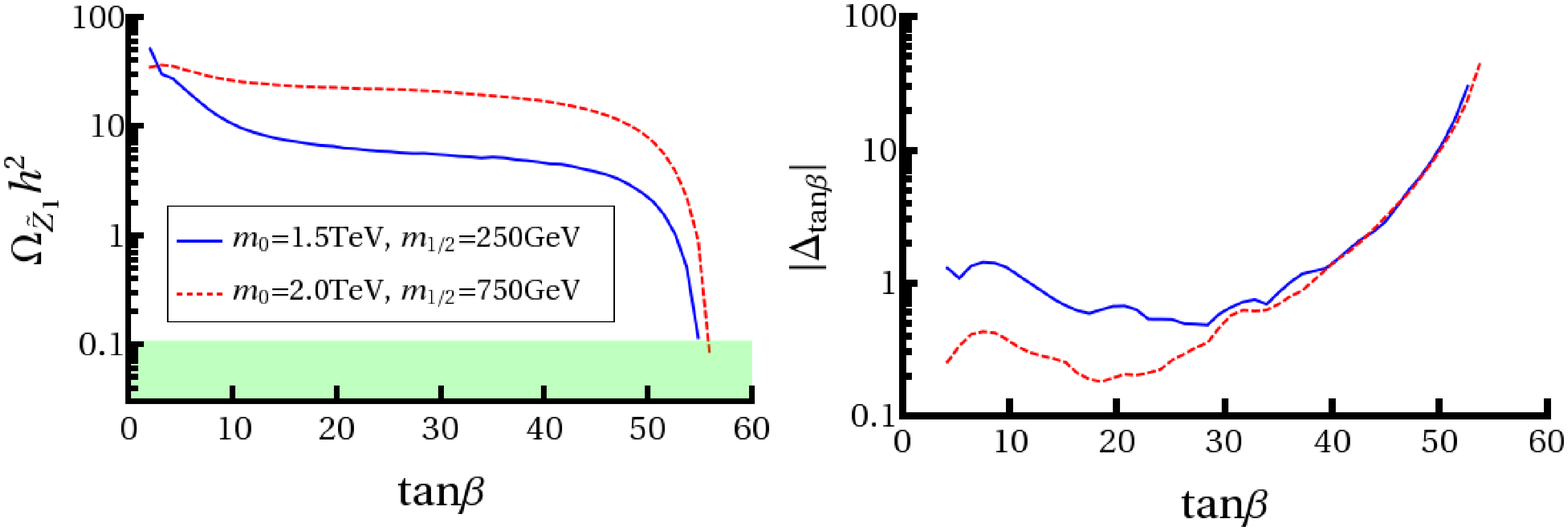}
\caption{Plot of {\it a}). $\Omega_{\tz_1}h^2$ and {\it b}). $|\Delta_{\tan\beta}|$ 
versus $\tan\beta$ for two slices out of mSUGRA parameter space: 
1. $m_0=1.5$ TeV and $m_{1/2}=250$ GeV (blue curves) and 2. $m_0=2$ TeV and 
$m_{1/2}=750$ GeV (red dashed curve), for $A_0=0$ and $\mu >0$.
}\label{fig:tanb}}

In Fig. \ref{fig:tanbplane}, we show the value of $\Delta$ including contributions from
variation in $m_0$, $m_{1/2}$ {\it and} $\tan\beta$, for the large values of
{\it a}). $\tan\beta =53$ and {\it b}). $\tan\beta =55$. 
Here, over essentially all of parameter space, the value of $\Delta$ has increased to much larger
values than those for $\Delta_{\tz_1}$ as shown in Fig's \ref{fig:ino53} and \ref{fig:ino55}. 
Thus, inclusion of $\tan\beta$ in the 
fine-tuning calculation shows that large values of $\tan\beta \agt 50$ result in
large fine-tuning of the relic density.
\FIGURE[t]{
\includegraphics[width=14cm]{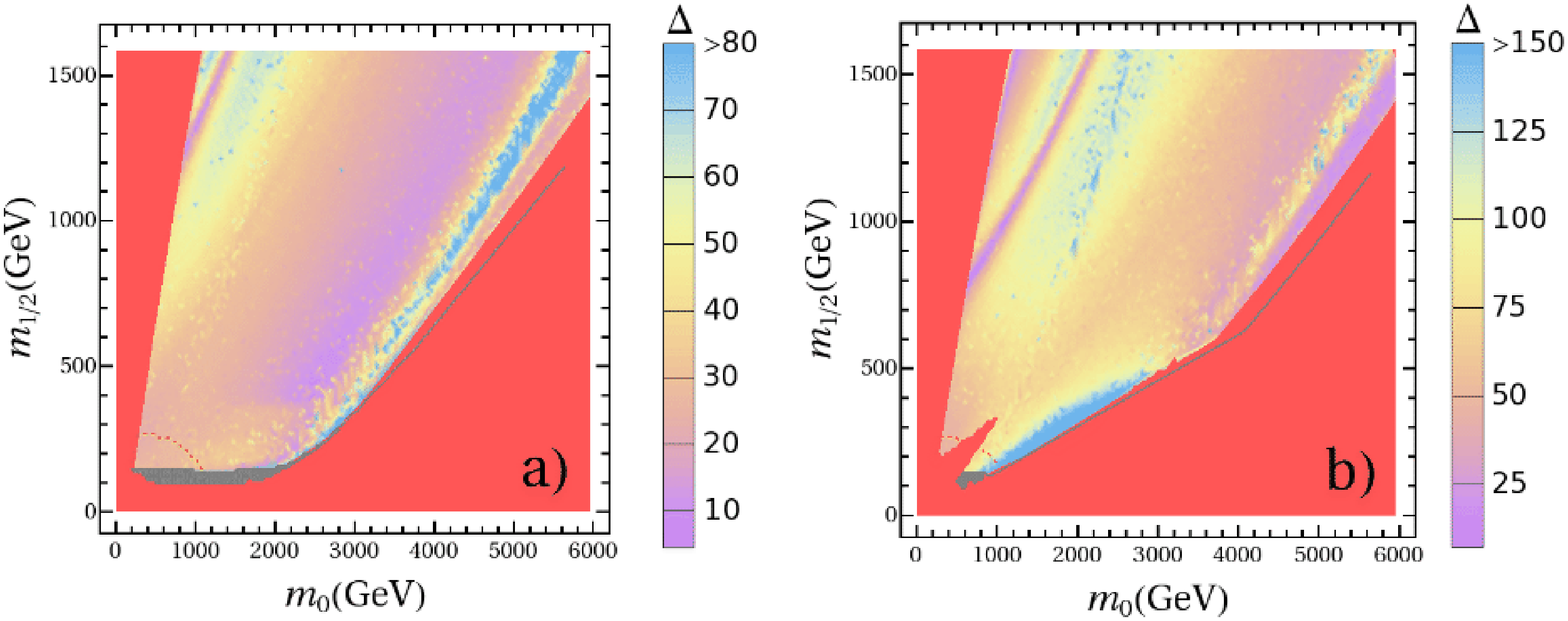}
\caption{Plot of $\Delta$ in the $m_0\ vs.\ m_{1/2}$ plane including variation of 
$m_0$, $m_{1/2}$ and $\tan\beta$ for $A_0=0$, $\mu >0$ and
{\it a}). $\tan\beta =53$ and {\it b}). $\tan\beta =55$.
}\label{fig:tanbplane}}

\section{The gravitino problem, leptogenesis, and the re-heat temperature}
\label{sec:inoprob}

In this section, we review the gravitino problem, baryogenesis via
leptogenesis, and production of mixed axion/axino dark matter in the 
early universe. The reader who is familiar with these issues may proceed directly to Sec. \ref{sec:axDM};
others may wish to follow the brief treatment given here and in Ref. \cite{axdm}.

\subsection{The gravitino problem}

In supergravity models, supersymmetry is broken via the superHiggs mechanism.
The common scenario is to postulate the existence of a hidden sector which
is uncoupled to the MSSM sector except via gravity. The superpotential of the 
hidden sector is chosen such that supergravity is broken, which causes 
the gravitino (which serves as the gauge particle for the superHiggs mechanism)
to develop a mass $m_{3/2}\sim m^2/M_{Pl}\sim m_{weak}$. Here, $m$ is a
hidden sector parameter assumed to be of order $10^{11}$ GeV.\footnote{
In Ref. \cite{kimnilles}, a link is suggested between 
hidden sector parameters and the PQ breaking scale $f_a$.} 
In addition to a mass
for the gravitino, SSB masses of order $m_{weak}$ are generated for
all scalar, gaugino, trilinear and bilinear SSB terms.
Here, we will assume that $m_{3/2}$ is larger than the lightest
MSSM mass eigenstate, so that the gravitino essentially decouples from
all collider phenomenology.

In all SUGRA scenarios, a potential problem arises for weak-scale
gravitinos: the gravitino problem. In this case, gravitinos $\tG$ can be
produced thermally in the early universe (even though the gravitinos
are too weakly coupled to be in thermal equilibrium) at a rate which
depends on the re-heat temperature $T_R$ of the universe. The produced $\tG$
can then decay to various sparticle-particle combinations, with a long 
lifetime of order $1-10^5$ sec (due to the Planck suppressed 
gravitino coupling constant). The late gravitino decays occur
during or after BBN, and their energy injection
into the cosmic soup threatens to destroy the successful BBN
predictions of the light element abundances.
The precise constraints of BBN on the gravitino mass and $T_R$ are
presented recently in Ref. \cite{kohri}. One way to avoid the  
gravitino problem in the case where $m_{3/2}\alt 5$ TeV is to maintain 
a value of $T_R\alt 10^5$ GeV. Such a low value of $T_R$ rules out many
attractive baryogenesis mechanisms, and so here instead we assume
that $m_{3/2}\agt 5$ TeV. In this case, the $\tG$ is so heavy that its
lifetime is of order 1 sec or less, and the $\tG$ decays near the onset of BBN.
In this case, values of $T_R$ as large as $10^9$ GeV are allowed.

In the simplest SUGRA models, one typically finds $m_0= m_{3/2}$.
For more general SUGRA models, the scalar masses are in general
non-degenerate and only of order $m_{3/2}$\cite{sugmasses}. 
Here for simplicity,
we will assume degeneracy of scalar masses, but with $m_0\ll m_{3/2}$.

\subsubsection{Leptogenesis}

One possible baryogenesis mechanism that requires relatively
low $T_R\sim m_{weak}$ is electroweak baryogenesis. However,
calculations of successful electroweak baryogenesis within 
the MSSM context seem to require sparticle mass spectra
with $m_h\alt 120$ GeV, and $m_{\tst_1}\alt 125$ GeV\cite{cnqw}. 
The latter requirement
is difficult (though not impossible) to achieve in the MSSM, and
is also partially excluded by collider searches for light
top squarks\cite{tev+stop}. 
We will not consider this possibility further.

An alternative attractive mechanism-- especially in light of recent
evidence for neutrino mass-- is thermal leptogenesis\cite{leptog}. 
In this scenario,
heavy right-handed neutrino states $N_i$ ($i=1-3$) 
decay asymmetrically to leptons
versus anti-leptons in the early universe. 
The lepton-antilepton asymmetry is converted to a baryon-antibaryon 
asymmetry via sphaleron effects. The measured baryon abundance can be achieved
provided the re-heat temperature $T_R$ exceeds $\sim 10^9$ GeV\cite{buchm}. 
The high $T_R$ value needed here apparently puts this mechanism into 
conflict with the gravitino problem in SUGRA theories.

A related leptogenesis mechanism called non-thermal leptogenesis
invokes an alternative to thermal production of heavy neutrinos 
in the early universe. 
In non-thermal leptogenesis\cite{NTlepto}, it is possible to have lower
re-heat temperatures, since the $N_i$ may be generated via inflaton decay.
The Boltzmann equations for the $B-L$ asymmetry have been solved numerically 
in Ref. \cite{imy}.
The $B-L$ asymmetry is then converted to a baryon asymmetry via sphaleron 
effects as usual.
The baryon-to-entropy ratio is calculated in \cite{imy}, where it is found
\be
\frac{n_B}{s}\simeq 8.2\times 10^{-11}\times \left(\frac{T_R}{10^6\ {\rm GeV}}\right)
\left(\frac{2M_{N_1}}{m_\phi}\right) \left(\frac{m_{\nu_3}}{0.05\ {\rm eV}}\right) \delta_{eff} ,
\ee
where $m_\phi$ is the inflaton mass and $\delta_{eff}$ is an effective $CP$ violating phase
which may be of order 1.
Comparing calculation with data (the measured value of $n_B/s\simeq 0.9\times 10^{-10}$), 
a lower bound $T_R\agt 10^6$ GeV may be 
inferred for viable non-thermal leptogenesis via inflaton decay.

A fourth mechanism for baryogenesis is Affleck-Dine\cite{ad} 
leptogenesis\cite{my}. In this approach, a flat direction 
$\phi_i =(2H\ell_i)^{1/2}$ is identified in the scalar potential,
which may have a large field value in the early universe. When the 
expansion rate becomes comparable to the SSB terms, the field oscillates,
and since the field carries lepton number, coherent oscillations about
the potential minimum will develop a lepton number asymmetry. The lepton
number asymmetry is then converted to a baryon number asymmetry by sphalerons
as usual. Detailed calculations\cite{my} find that the
baryon-to-entropy ratio is given by
\be
\frac{n_B}{s}\simeq\frac{1}{23}\frac{|\langle H\rangle |^2  T_R}
{m_\nu M_{Pl}^2}
\ee
where $\langle H\rangle$ is the Higgs field vev, $m_\nu$ is the mass of the
lightest neutrino and $M_{Pl}$ is the Planck scale. To obtain the observed 
value of $n_B/s$, values of $T_R\sim 10^6-10^8$ are allowed for
$m_{\nu}\sim 10^{-9}-10^{-7}$ eV.

Thus, to maintain accord with either non-thermal or Affleck-Dine
leptogenesis, along with constraints from the gravitino problem, we will
aim for $a\tilde{a}$DM scenarios with $T_R\sim  10^6-10^8$ GeV.

\subsection{Mixed axion/axino dark matter}

\subsubsection{Relic axions}

Axions can be produced via various mechanisms
in the early universe. Since their lifetime 
(they decay via $a\to\gamma\gamma$) turns out to
be longer than the age of the universe, 
they can be a good candidate for dark matter.
As we will be concerned here with re-heat temperatures 
$T_R\alt 10^9\ {\rm GeV }<f_a/N$ 
(to avoid overproducing gravitinos in the early universe), 
the axion production mechanism relevant
for us here is just one: 
production via vacuum mis-alignment\cite{absik}. In this mechanism, the axion field
$a(x)$ can have any value $\sim f_a$ at temperatures $T\gg \Lambda_{QCD}$. As the temperature
of the universe drops,
the potential turns on, and the axion field oscillates and settles to its minimum 
at $-\bar{\theta} f_a/N$ (where $\bar{\theta}=\theta +arg(det\ m_q)$,
$\theta$ is the fundamental strong $CP$ violating Lagrangian parameter and $m_q$ is
the quark mass matrix).
The difference in axion field before and after potential turn-on corresponds to
the vacuum mis-alignment: it produces an axion number density
\be
n_a(t)\sim {\frac{1}{2}}m_a(t)\langle a^2(t)\rangle ,
\ee
where $t$ is  the time near the QCD phase transition.
Relating the number density to the entropy density allows one to determine the
axion relic density today\cite{absik}:
\be
\Omega_a h^2\simeq {\frac{1}{4}}\left(\frac{6\times 10^{-6}\ {\rm eV}}{m_a}\right)^{7/6}\theta_i^2
\simeq\frac{1}{4}\left(\frac{f_a/N}{10^{12}\ \mathrm{GeV}}\right)^{7/6}\theta_i^2 ,
\label{eq:axrd}
\ee
where $\theta_i$ is the initial vacuum mis-alignment angle, with $-\pi\alt \theta_i\alt\pi$.               
An error estimate of the axion relic density from vacuum mis-alignment is
plus-or-minus a factor of three.
Axions produced via vacuum mis-alignment would constititute {\it cold} dark matter.
However, in the event that $\langle a^2(t)\rangle$ is
inadvertently small, then much lower values of axion relic density could be allowed.
Additional entropy production at $t>t_{QCD}$ can also lower the axion relic abundance.
Taking the value of Eq.~(\ref{eq:axrd}) literally, along with $\theta_i\simeq 1$,
and comparing to the WMAP5 measured abundance of CDM in the universe,
one gets an upper bound $f_a/N\alt 5\times 10^{11}$ GeV, or a lower bound
$m_a\agt 10^{-5}$ eV. If we take the axion
relic density a factor of three lower, then the bounds change
to $f_a/N \alt 1.2\times 10^{12}$ GeV, and $m_a\agt 4\times 10^{-6}$ eV.

\subsubsection{Axinos from neutralino decay}

If the $\ta$ is the lightest SUSY particle, then the $\tz_1$ will no longer
be stable, and can decay via $\tz_1\to \ta\gamma$.
The relic abundance of axinos from neutralino decay
(non-thermal production, or $NTP$) is given simply by
\be
\Omega_{\ta}^{NTP}h^2 =\frac{m_{\ta}}{m_{\tz_1}}\Omega_{\tz_1}h^2 ,
\label{eq:Oh2_NTP}
\ee
since in this case the axinos inherit the thermally produced
neutralino number density.
The neutralino-to-axino decay offers a mechanism to shed
large factors of relic density. For a case where $m_{\tz_1}\sim 100$
GeV and $\Omega_{\tz_1}h^2\sim 10$ (as can occur in the mSUGRA model
at large $m_0$ values)
an axino mass of less than 1 GeV reduces the DM abundance to below
WMAP-measured levels.

The lifetime for these decays has been calculated,
and it is typically in the range of $\tau (\tz_1\to \ta\gamma )\sim 0.01-1$ sec\cite{ckkr}.
The photon energy injection from $\tz_1\to\ta\gamma$ decay
into the cosmic soup occurs typically before
BBN, thus avoiding the constraints that plague the case of a gravitino LSP\cite{kohri}.
The axino DM arising from neutralino decay is generally
considered warm or even hot dark matter for cases with
$m_{\ta}\alt 1-10$ GeV\cite{jlm}.
Thus, in the mSUGRA scenario considered here, where $m_{\ta}\alt 1-10$ GeV, we usually get {\it warm} axino DM from neutralino decay.

\subsubsection{Thermal production of axinos}

Even though axinos may not be in thermal equilibrium in the early universe, 
they can still be produced thermally via scattering and decay processes in the cosmic soup.
The axino thermally produced (TP) relic abundance has been
calculated in Ref. \cite{ckkr,steffen}, and is given in Ref. \cite{steffen} using
hard thermal loop resummation as
\be
\Omega_{\ta}^{TP}h^2\simeq 5.5 g_s^6\ln\left(\frac{1.211}{g_s}\right)
\left(\frac{10^{11}\ {\rm GeV}}{f_a/N}\right)^2
\left(\frac{m_{\ta}}{0.1\ {\rm GeV}}\right)
\left(\frac{T_R}{10^4\ {\rm GeV}}\right)
\label{eq:Oh2_TP}
\ee
where $g_s$ is the strong coupling evaluated at $Q=T_R$ and $N$ is the
model dependent color anomaly of the PQ symmetry, of order 1.
For reference, we take $g_s(T_R=10^6\ {\rm GeV})=0.932$ 
(as given by Isajet 2-loop $g_s$ evolution in mSUGRA), with $g_s$ at 
other values of $T_R$ given by the 1-loop MSSM running value.
The thermally produced axinos qualify as {\it cold} dark matter as long as
$m_{\ta}\agt 0.1$ MeV\cite{ckkr,steffen}.

\section{Fine-tuning in mSUGRA with mixed axion/axino CDM}
\label{sec:axDM}

In this section, we calculate the fine-tuning parameter for the
dark matter relic density in models with mixed $a\tilde{a}$DM: $\Delta_{a\ta}$.
Contributions to $\Delta_{a\ta}$ are calculated from both the axion relic density and the 
thermally produced axino relic density. 
We do not include the non-thermally produced axino relic density as it makes a tiny
contribution to the total for the values of $m_{\ta}\sim 1$ MeV considered here
(see Figs. 2 and 3 of Ref. \cite{axdm}). We take the total relic density to be
\begin{align}
\Omega_{a\tilde{a}}h^2&=\Omega_ah^2+\Omega^{TP}_{\tilde{a}}h^2\\
\label{eq:fulldel}&=\frac{1}{4}\left(\frac{f_a/N}{10^{12}\ \mathrm{GeV}}\right)^{7/6}\theta_i^2
+5.5g^6_s\ln{\left(\frac{1.211}{g_s}\right)}\left(\frac{10^{11}\ \mathrm{GeV}}{f_a/N}\right)^2\left(\frac{m_{\tilde{a}}}{0.1\ \mathrm{GeV}}\right)\left(\frac{T_R}{10^4\ \mathrm{GeV}}\right)
\end{align}
and calculate the total $\Delta_{a\ta}$ exactly by differentiating \eqref{eq:fulldel} with respect to 
$f_a/N$, $T_R$ and $m_{\tilde{a}}$.\footnote{Here, one objection may be that the value of $T_R$
does not appear as an explicit Lagrangian parameter. However, in the standard inflationary cosmology,
the reheat temperature is related to the inflaton decay width via
$T_R\simeq(3/\pi^3)^{1/4}g_*^{-1/4}(M_{Pl}\Gamma_\phi )^{1/2}$\cite{kolbturner}, 
where $\Gamma_\phi$ depends on the inflaton mass and couplings to matter. 
In this case, a detailed model including the inflaton field $\phi$ 
would provide $T_R$ in terms of inflaton Lagrangian parameters. We do not wish to bring such model-dependence
into our calculations, so instead just adopt the value of $T_R$ as a fundamental parameter. Also, 
the value of $m_{\ta}$ will appear as a Lagrangian parameter in the weak scale effective Lagrangian, after the
effects of SUSY breaking and PQ breaking are taken into account.
} 
We find:
\bea
\Delta_{T_R}&=\frac{T_R}{\Omega_{a\tilde{a}}h^2}\frac{\partial \Omega_{a\tilde{a}}h^2}{\partial T_R}=\frac{T_R}{\Omega_{a\tilde{a}}h^2}5.5g^6_s\ln{\left(\frac{1.211}{g_s}\right)}\left(\frac{10^{11}\ \mathrm{GeV}}{f_a/N}\right)^2\left(\frac{m_{\tilde{a}}}{0.1\ \mathrm{GeV}}\right)\left(\frac{1}{10^4\ \mathrm{GeV}}\right)\\
\Delta_{m_{\tilde{a}}}&=\frac{m_{\tilde{a}}}{\Omega_{a\tilde{a}}h^2}\frac{\partial \Omega_{a\tilde{a}}h^2}{\partial m_{\tilde{a}}}=\frac{m_{\tilde{a}}}{\Omega_{a\tilde{a}}h^2}5.5g^6_s\ln{\left(\frac{1.211}{g_s}\right)}\left(\frac{10^{11}\ \mathrm{GeV}}{f_a/N}\right)^2\left(\frac{1}{0.1\ \mathrm{GeV}}\right)\left(\frac{T_R}{10^4\ \mathrm{GeV}}\right)
\eea
and
\begin{align}
\Delta_{f_a/N}&=\frac{f_a/N}{\Omega_{a\tilde{a}}h^2}\frac{\partial \Omega_{a\tilde{a}}h^2}{\partial f_a/N}\\\begin{split}
&=\frac{f_a/N}{\Omega_{a\tilde{a}}h^2}\left[\frac{7}{24}\left(\frac{1}{10^{12}\ \mathrm{GeV}}\right)^{7/6}\left(f_a/N\right)^{1/6}\theta_i^2\right.\\
&\qquad\qquad-\left. 11g^6_s\ln{\left(\frac{1.211}{g_s}\right)}\left(10^{11}\ \mathrm{GeV}\right)^2\left(\frac{1}{f_a/N}\right)^3\left(\frac{m_{\tilde{a}}}{0.1\ \mathrm{GeV}}\right)\left(\frac{T_R}{10^4\ \mathrm{GeV}}\right)\right] ,
\end{split}\end{align}
and
\be
\Delta_{\theta_i}=\frac{\theta_i}{\Omega_{a\tilde{a}}h^2}\frac{\partial \Omega_{a\tilde{a}}h^2}{\partial \theta_i}=2\frac{\Omega_ah^2}{\Omega_{a\tilde{a}}h^2}.
\ee
The total fine tuning parameter is then given by
\begin{equation}
\Delta_{a\ta} =\sqrt{\Delta_{T_R}^2+\Delta_{m_{\tilde{a}}}^2+\Delta_{f_a/N}^2+\Delta_{\theta_i}^2} .
\end{equation}

We plot our first results in the $f_a/N\ vs.\ T_R$ plane, keeping $m_{\ta}$ fixed at 1 MeV:
see Fig. \ref{fig:max-3}.
In frame {\it a})., we show contours of $\Omega_{a\ta}h^2$. The green region gives
$\Omega_{a\ta}<0.11$, and so is consistent with WMAP. In frame {\it b})., we show regions
of fine-tuning $\Delta_{a\ta}$. The scale is shown on the right edge of the plot. Note in this case
the entire plane has $\Delta_{a\ta}<2.5$, so there is very little fine-tuning across the entire
plane of parameter space. The left region, color-coded dark blue, is the region of 
dominantly thermally produced {\it axino CDM}, whilst the right-most region, color-coded lighter blue,
is dominantly {\it axion CDM}. In this region, the fine-tuning parameter $\Delta_{a\ta}\simeq 2.3$.
The intermediate region, shaded by yellow and purple bands, is the region of mixed $a\tilde{a}$CDM: 
the purple band has very low fine-tuning, with $\Delta_{a\ta}<1.4$. The contour
where mixed $a\tilde{a}$CDM saturate the WMAP measured value is shown by the green dashed line.
The region where the highest values of $T_R$ are found coincide with the region of lowest fine-tuning, 
with a nearly equal mix of axion and axino CDM.
\FIGURE[t]{
\includegraphics[width=14cm]{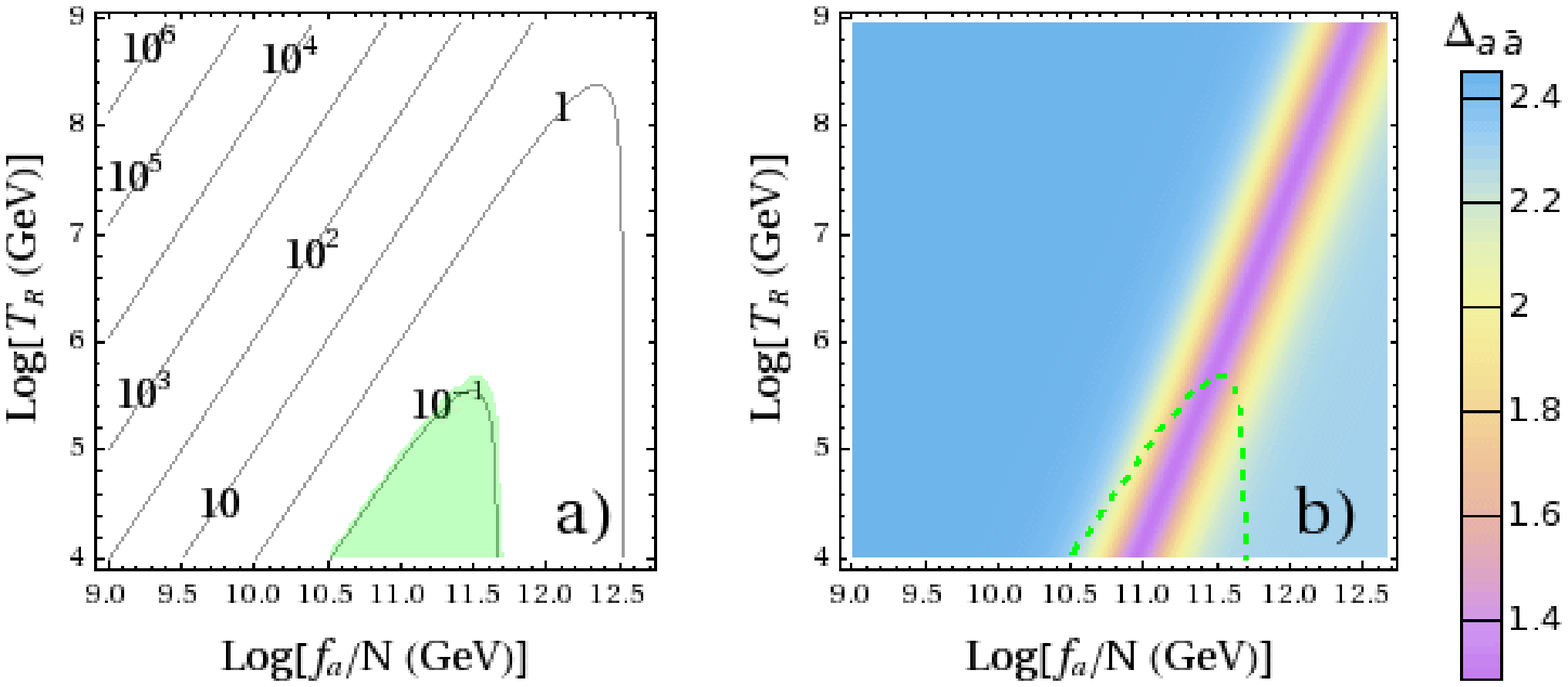}
\caption{Plot of {\it a}). contours of axion/axino relic density
$\Omega_{a\ta}h^2$ in the $f_a/N\ vs.\ T_R$ plane
and {\it b}). regions of fine-tuning $\Delta_{a\ta}$ for
fixed axino mass $m_{\ta}=1$ MeV.
The green region in {\it a}). has $\Omega_{a\ta}h^2\le 0.11$. The green dashed line
in frame {\it b}). is where $\Omega_{a\ta}h^2= 0.11$.
}\label{fig:max-3}}

To gain additional perspective, in Fig. \ref{fig:3da} we show the mixed axion/axino relic density 
as a 3-d plot in the $f_a/N\ vs.\ T_R$ plane for $m_{\ta}=1$ MeV. The level of fine-tuning, corresponding
to the slope of the surface, is rather low throughout, since there are no regions with a steep slope.
The fine-tuning is minimal along the trough running through the right-center of the plot. 
\FIGURE[t]{
\includegraphics[width=14cm]{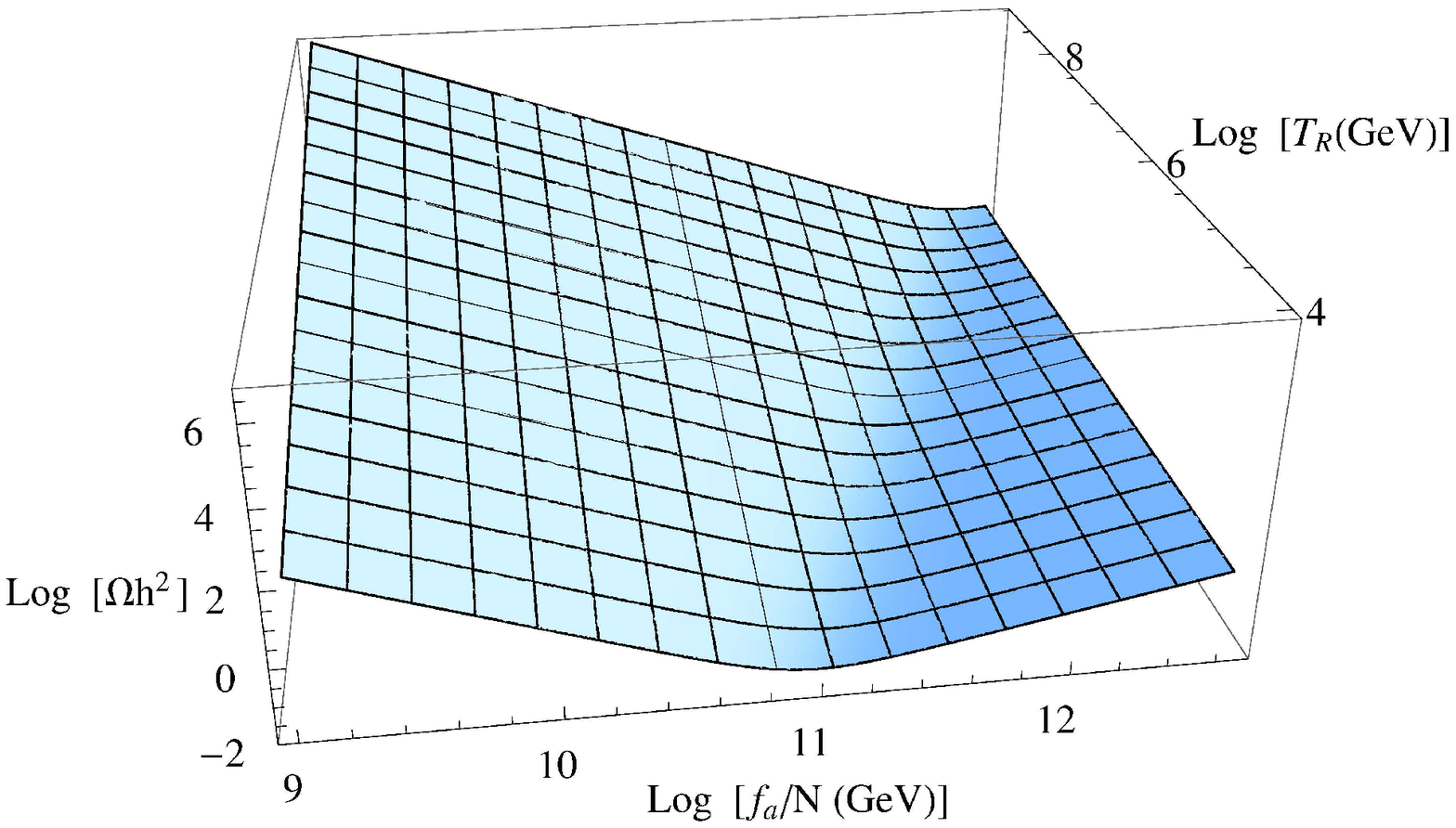}
\caption{A 3-d plot of axion/axino relic density in the $f_a/N\ vs.\ T_R$ 
plane of the $a\ta$ augmented mSUGRA model.
}\label{fig:3da}}

To better understand the situation with mixed $a\tilde{a}$CDM, we show in Fig. \ref{fig:slice_max-3}
a slice of our Fig. \ref{fig:max-3} with constant $T_R=10^5$ GeV.
In frame {\it a})., we see that the value of $\Omega_{a\ta}h^2$ initially drops as
$f_a/N$ increases. This is in the region of dominant axino CDM, and increasing $f_a/N$
decreases the axino coupling strength, and hence suppresses its thermal production in the early universe.
As $f_a/N$ increases further, the relic abundance of axions steadily increases, until
around $f_a/N\sim 2\times 10^{11}$ GeV there is an upswing in the relic abundance. This is the 
stable fine-tuning region since small fluctuations of parameters about this point do not substantially alter the
axino/axino relic density. The fine-tuning parameter $\Delta_{a\ta}$ is shown in frame {\it b})..
Here, we see that the fine-tuning is slightly high in the region of mainly axino CDM, with low
$f_a/N$, but reaches a minimum at the point of equal admixture. The value of $\Delta_{a\ta}$ doesn't extend
all the way to zero, in spite of the zero slope shown, because $\Delta_{a\ta}$ still varies with 
$m_{\ta}$, $T_R$ and $\theta_i$. The fine-tuning parameter increases to the analytic value of $2.3$ as
$f_a/N$ increases further, into the region of mainly axion CDM.
\FIGURE[t]{
\includegraphics[width=14cm]{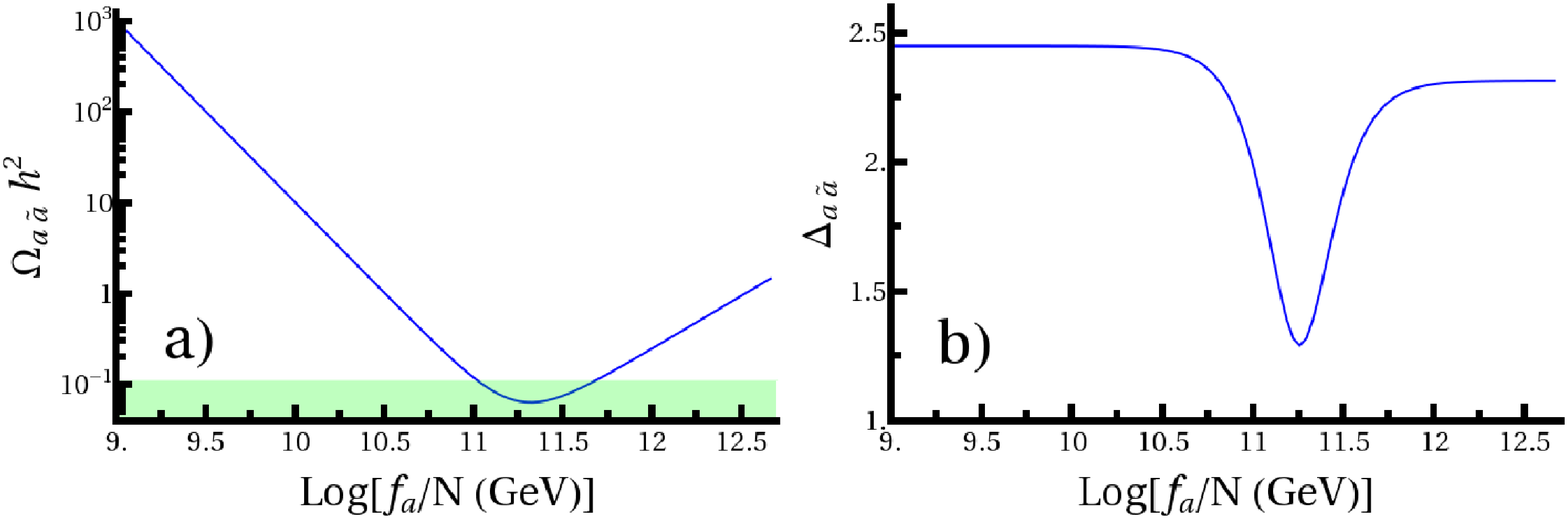}
\caption{Plot of {\it a}). axion/axino relic density
$\Omega_{a\ta}h^2$ 
and {\it b}). fine-tuning parameter $\Delta_{a\ta}$ 
versus $f_a/N$ for fixed axino mass $m_{\ta}=1$ MeV and fixed re-heat temperature
$T_R=10^5$ GeV.
}\label{fig:slice_max-3}}

A similar plot to Fig. \ref{fig:max-3} is shown in Fig. \ref{fig:max-4}, but in this case taking
$m_{\ta}=0.1$ MeV. Note that this value yields the approximate dividing line given in Refs. \cite{ckkr,steffen}
below which the thermally produced axinos would be mainly warm DM instead of cold DM. In any case, in frame {\it a}).,
we see that the WMAP allowed region has expanded, and now values of $T_R$ as high as $5\times10^{6}$ GeV are allowed, 
making the scenario consistent with at least non-thermal leptogenesis. The region of maximal $T_R$ also
coincides with the region of least fine-tuning, with a roughly equal admixture of axion and thermally produced axino DM.
%
\FIGURE[t]{
\includegraphics[width=14cm]{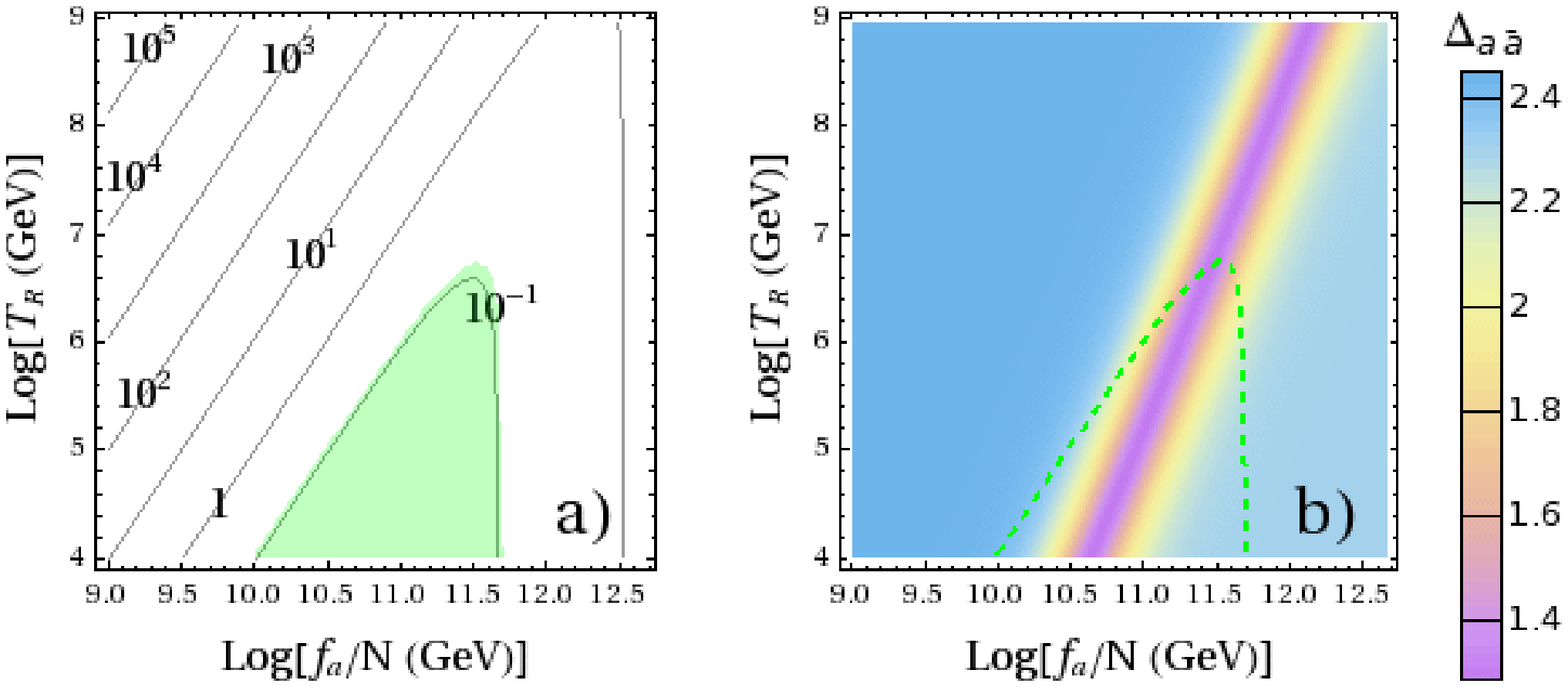}
\caption{Plot of {\it a}). contours of axion/axino relic density
$\Omega_{a\ta}h^2$ in the $f_a/N\ vs.\ T_R$ plane
and {\it b}). regions of fine-tuning $\Delta_{a\ta}$ for
fixed axino mass $m_{\ta}=0.1$ MeV.
The green region in {\it a}). has $\Omega_{a\ta}h^2\le 0.11$. The green dashed line
in frame {\it b}). is where $\Omega_{a\ta}h^2= 0.11$.
}\label{fig:max-4}}

In Fig. \ref{fig:ata_fafix1} we show the contours of relic density $\Omega_{a\ta}h^2$ in the 
$m_{\ta}\ vs.\ T_R$ plane for fixed value of $f_a/N=4.88\times 10^{11}$ GeV. The large value of
$f_a/N$ yields a scenario with mainly axion CDM when the WMAP measured abundance is saturated.
The green shaded region in frame {\it a}). is WMAP-allowed. The red dashed line shows the approximate
dividing line between warm and cold thermally produced axinos. In this case, the demarcation line is
largely irrelevant, since if $\Omega_{a\ta}h^2\simeq 0.11$, almost all the DM is composed of cold axions,
and a tiny admixture of warm axinos would be allowed. 
In frame {\it b})., we show the regions of fine-tuning $\Delta_{a\ta}$. Since the WMAP-allowed region
coincides with mainly axion CDM, the fine-tuning along the green dashed line is always low:
$\Delta_{a\ta}\sim 2.3$.
Note that in the scenario with mainly axion CDM, the value of $T_R$ can easily reach to well over 
$10^7$ GeV, allowing for non-thermal or Affleck-Dine leptogenesis.
\FIGURE[t]{
\includegraphics[width=14cm]{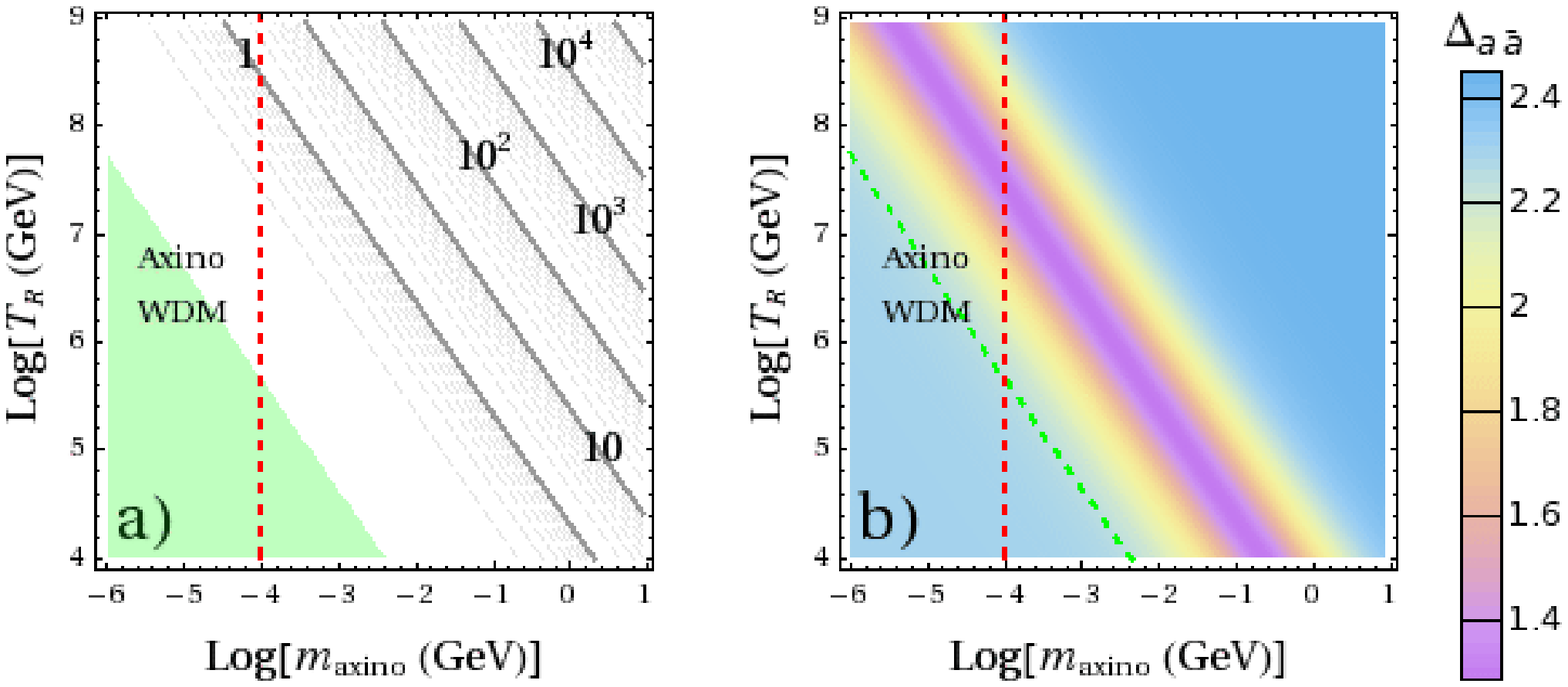}
\caption{Plot of {\it a}). contours of axion/axino relic density
$\Omega_{a\ta}h^2$ in the $m_{\ta}\ vs.\ T_R$ plane
and {\it b}). regions of fine-tuning $\Delta_{a\ta}$ for
fixed $f_a/N=4.88\times 10^{11}$ GeV (which gives mainly axion CDM along the line
of $\Omega_{a\ta}=0.11$).
The green region in {\it a}). has $\Omega_{a\ta}h^2\le 0.11$. The green dashed line
in frame {\it b}). is where $\Omega_{a\ta}h^2= 0.11$.
The region to the left of red-dashed line gives thermally produced {\it warm} 
axino dark matter.
}\label{fig:ata_fafix1}}

In Fig. \ref{fig:ata_fafix2} we show the $m_{\ta}\ vs.\ T_R$ plane for $f_a/N=3\times 10^{11}$ GeV, 
which gives roughly an equal admixture of axion and thermally produced axino DM. In this case, 
the value of $T_R$ reaches beyond $10^8$ GeV, although for very low values of $m_{\ta}$ where
it is expected that the axino will be warm DM. For this scenario, it is unclear how much mixture of
warm and cold dark matter is cosmologically allowed. To answer the question, the velocity profile of the
warm axinos would have to be fed into $n-body$ simulations of large scale structure formation, 
to see how well such a mixed warm/cold DM scenario fits the data. At present, we are unaware of such studies.
In frame {\it b})., we see that the line of WMAP-saturated abundance lies nearly on top of the region of lowest
fine-tuning, with $\Delta_{a\ta}<1.4$. 
\FIGURE[t]{
\includegraphics[width=14cm]{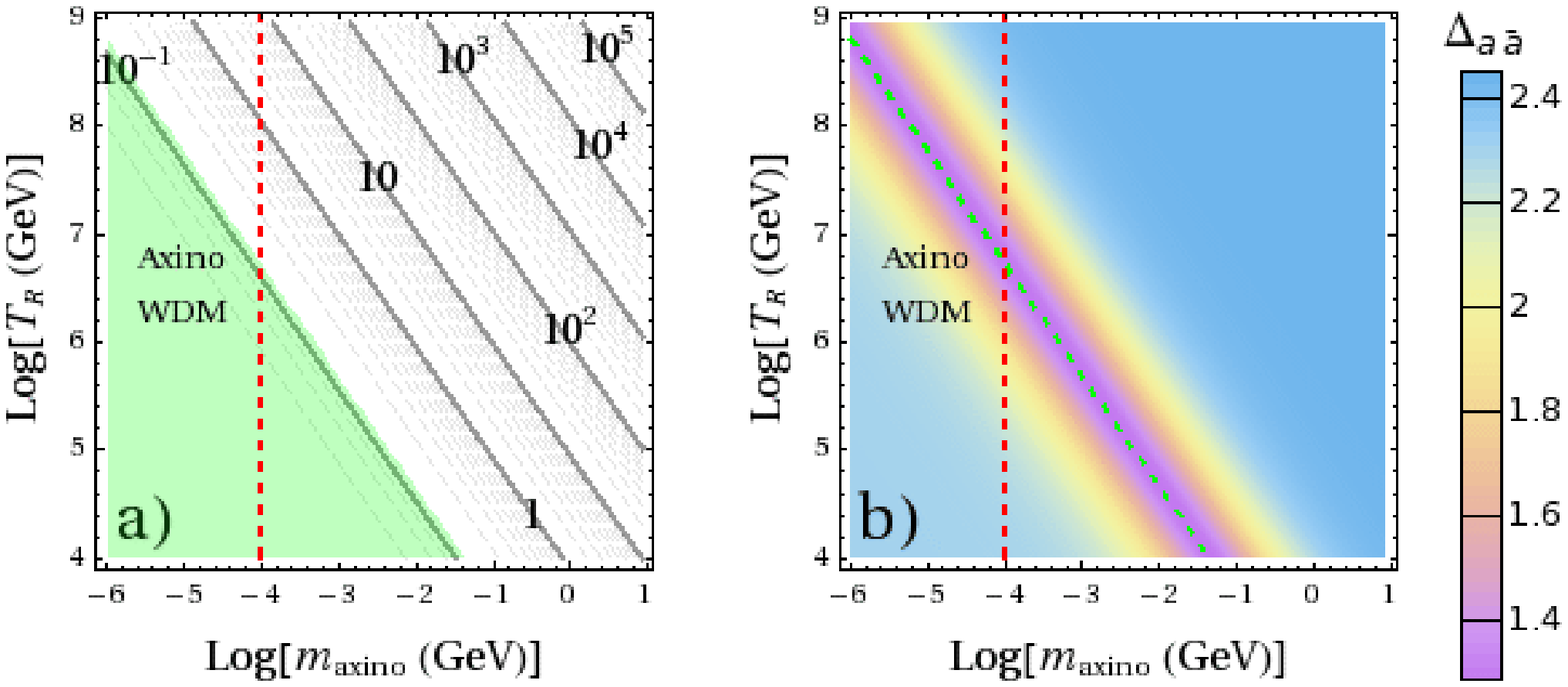}
\caption{Plot of {\it a}). contours of axion/axino relic density
$\Omega_{a\ta}h^2$ in the $m_{\ta}\ vs.\ T_R$ plane
and {\it b}). regions of fine-tuning $\Delta_{a\ta}$ for
fixed $f_a/N=3\times 10^{11}$ GeV (which gives a 50-50 mix of axion/axino DM along the line
of $\Omega_{a\ta}=0.11$).
The green region in {\it a}). has $\Omega_{a\ta}h^2\le 0.11$. The green dashed line
in frame {\it b}). is where $\Omega_{a\ta}h^2= 0.11$.
The region to the left of red-dashed line gives thermally produced {\it warm} 
axino dark matter.
}\label{fig:ata_fafix2}}

In Fig. \ref{fig:ata_fafix3}, we show again the $m_{\ta}\ vs.\ T_R$ plane, but this time for
$f_a/N=1\times 10^{11}$ GeV, which gives mainly {\it thermally produced axino} DM. In this case
the region to the left of the red-dashed line should likely be disallowed, since the dominant
form of DM will be warm, rather than cold. The region to the right of the $m_{\ta}=0.1$ MeV line,
in the WMAP-allowed region, only allows for $T_R$ to reach a max of $10^6$ GeV. Furthermore,
from frame {\it b})., we see that the fine-tuning parameter in this case for the
WMAP-saturated region along the green dashed line is somewhat higher, reaching $\Delta_{a\ta}\sim 2$.
\FIGURE[t]{
\includegraphics[width=14cm]{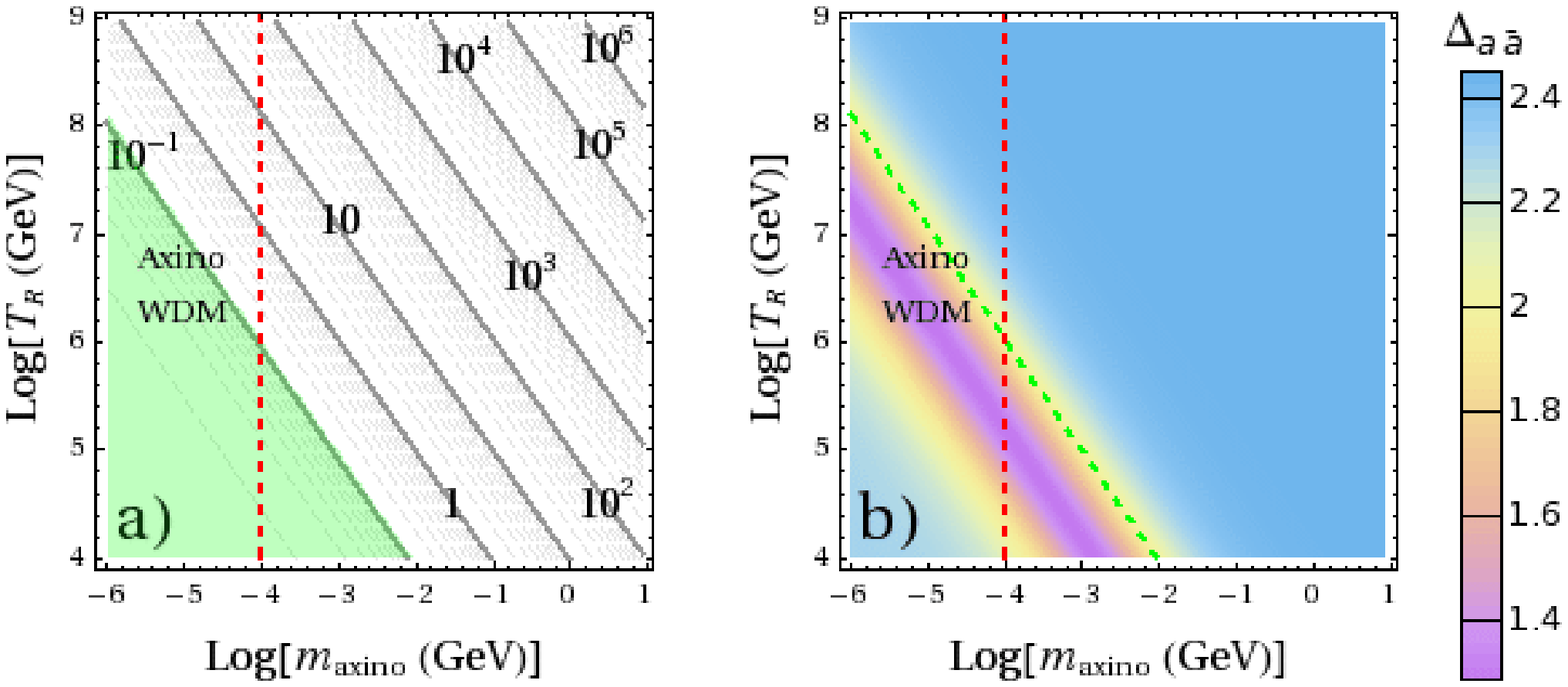}
\caption{Plot of {\it a}). contours of axion/axino relic density
$\Omega_{a\ta}h^2$ in the $m_{\ta}\ vs.\ T_R$ plane
and {\it b}). regions of fine-tuning $\Delta_{a\ta}$ for
fixed $f_a/N=1\times 10^{11}$ GeV (which gives mainly axino DM along the line
of $\Omega_{a\ta}=0.11$).
The green region in {\it a}). has $\Omega_{a\ta}h^2\le 0.11$. The green dashed line
in frame {\it b}). is where $\Omega_{a\ta}h^2= 0.11$.
The region to the left of red-dashed line gives thermally produced {\it warm} 
axino dark matter.
}\label{fig:ata_fafix3}}
%

\section{Summary and conclusions}
\label{sec:conclude}

In this paper, we have examined the fine-tuning associated 
with the relic density of dark matter in the minimal
supergravity model. We have calculated a measure of fine-tuning assuming two
scenarios for SUSY dark matter: 1. neutralino dark matter with
fine-tuning parameter $\Delta_{\tz_1}$, and 2. mixed axion/axino dark matter
with fine-tuning parameter $\Delta_{a\ta}$. 

In the case of neutralino dark matter, we find that the WMAP-allowed regions of
mSUGRA such as the stau co-annihilation region, the HB/FP region and the 
light Higgs $h$-resonance annihilation region, are all rather highly fine-tuned, 
especially the HB/FP region, where $\Delta_{\tz_1}$ ranges from 20-100. 
Only mild fine-tuning is found in the low $m_0$, low $m_{1/2}$ region 
where stau co-annihilation and bulk annihilation through $t$-channel slepton
exchange overlap. If one moves to large $\tan\beta\sim 50$, then larger
regions of parameter space which are consistent with WMAP occur. These large
$\tan\beta$ regions have modest fine-tuning versus $m_0$ and $m_{1/2}$, but very 
large fine-tuning versus $\tan\beta$.

If instead we assume that dark matter is composed of an axion/axino
admixture, rather than neutralinos, then we find that the relic density 
fine-tuning parameter is generically much lower: $\Delta_{a\ta}\sim 1.3-2.5$
throughout parameter space. 
Here, we have assumed the existence of a light axino with mass $m_{\ta}\sim$ keV-MeV.
Such a light axino opens up all of mSUGRA parameter space to being WMAP allowed, 
since now the neutralino decays via $\tz_1\to \ta\gamma$. If the DM is dominated by
thermally produced axinos, then the re-heat temperature $T_R$ is generally lower than
$10^6$ GeV unless the axinos are actually warm dark matter ($m_{\ta}\alt 100$ keV), 
so this scenario seems rather unlikely. However, if the PQ breaking scale $f_a/N$ is large, then
the DM can be either a nearly equal axion/axino admixture, in which case fine-tuning is lowest
($\Delta_{a\ta}\sim 1.3$), or a dominantly axion mixture (in which case
$\Delta_{a\ta}\sim 2.3$). Either scenario easily admits $T_R>10^6$ GeV, which can allow for
non-thermal leptogenesis to occur.

The consequences of the mixed $a\tilde{a}$CDM scenario for future dark matter searches is as follows.
For collider searches, we expect much the same collider signatures as in the mSUGRA model with
neutralino dark matter, since we assume the $\tz_1$ is the NLSP, and decays far outside the collider 
detectors. 
However, {\it all} of mSUGRA parameter space is now WMAP-allowed, instead of just the special
co-annihilation, HB/FP region and resonance annihilation regions. As shown in
Ref. \cite{axdm}, the regions of WMAP-allowed neutralino CDM yield the lowest
values of $T_R$, and so the stau co-annihilation, HB/FP region and $h$ resonance
annihilation regions are most dis-favored for the case of mixed $a\ta$CDM.

As far as WIMP searches go, in the mixed 
$a\tilde{a}$CDM scenario, we expect no positive signals if $m_{\tz_1}>m_{\ta}$. If 
$m_{\ta}>m_{\tz_1}$, then the $\tz_1$ would still be stable (assuming $R$-parity conservation) and WIMP
direct and indirect detection signals are still possible\cite{njp}. In the case of large axion relic abundance,
which appears to us to be the favored scenario, then a positive signal at relic axion search experiments
such as ADMX might be expected\cite{admx}, although solar axion searches are less likely to achieve positive results,
since large values of $f_a/N$ are favored, leading to small axion/axino couplings.

Our analysis has been based on the admittedly subjective basis of fine-tuning of the
relic density of dark matter relative to model input parameters. We note here that
the mSUGRA model already needs substantial fine-tuning in the electroweak sector in order to
accomodate the relatively light $Z$ boson mass in the face of limits on the soft 
SUSY breaking parameters\cite{ewft} (the little hierarchy problem). Our philosophy here is 
that less fine-tuning is better, and high fine-tuning in one sector is better than 
high fine-tuning in two sectors, {\it e.g.} electroweak and dark matter sectors.

While our analysis has been restricted to the mSUGRA SUSY model, one might ask 
how general our conclusions might be.
In SUSY models based on gravity mediation, with a neutralino LSP, the
DM relic density is {\it generically too high} unless some special mechanism is acting to
enhance the neutralino annihilation cross section in the early universe.\footnote{
Discussion on numerous different SUGRA models with non-universality has been explored in 
Ref. \cite{wtn}.}
For instance, in SUSY models with non-universality, instead of stau or stop co-annihilation, 
one might have sbottom or sneutrino co-annihilation, or
bino-wino co-annihilation: in any case, the mass gap between co-annihilating 
particles must be fine-tuned to obtain agreement with the measured dark matter abundance. 
In non-universal models with a {\it well-tempered neutralino}\cite{wtn}, where
the neutralino bino-higgsino or bino-wino composition is adjusted to fit the 
measured relic density, 
other parameters (Higgs soft masses, gaugino masses) must be fine-tuned to get just the
right ``tempering'', as occurs in the mSUGRA HB/FP region. 
In other models, Higgs soft mass terms can be adjusted to allow $2m_{\tz_1}$ to sit atop the 
$A$ resonance; but again, in this case, parameters must be
fine-tuned (unless $\tan\beta$ is large, which also occurs in mSUGRA). The case where the SUSY 
neutralino abundance is not fine-tuned has long been noted: it is where squarks and sleptons
are so light that $t$-channel annihilation channels are large. 
However, LEP2 search limits now essentially exclude all these regions.
Thus, although we restrict our analysis here to the mSUGRA model, we feel this model provides a sort of
microcosm for general SUSY models, in that it illustrates many of the 
features common to all SUSY models.

Our main conclusion is this. In the world HEP community, a tremendous effort is underway to explore
for WIMP cold dark matter, based partly on the view that SUSY models naturally 
give rise to the ``WIMP-miracle'', and an excellent WIMP candidate for CDM. 
We have shown here that at least for the
paradigm SUSY model-- mSUGRA-- usually a large overabundance of neutralino CDM is produced, unless one
lies along a region of very high fine-tuning, where a slight change in model parameters leads to a large
change in relic density: this equates to a high degree of relic density fine-tuning. Alternatively, if one
assumes the PQWW solution to the strong CP problem within SUSY models, and a very light axino with
$m_{\ta}$ of the order of MeV, then along with an elegant solution to the strong CP problem, one obtains a 
mixed axion/axino relic density with much less fine-tuning. Given our results, we would advocate that a 
much increased share of HEP resources be given to relic axion searches, where the global search effort 
has been much more limited.

\acknowledgments

We thank H. Summy for discussions.
This research was supported in part by the U.S. Department of Energy.
	
%


\begin{thebibliography}{99}
%
\bibitem{wmap5}
J.~Dunkley {\it et al.}  [WMAP Collaboration],
  Astrophys.\ J.\ Suppl.\  {\bf 180} (2009) 306
%
\bibitem{kcreview} For a recent review, 
see P. Sikivie, \hepph{0509198}; M. Turner, \prep{197}{1990}{67}; 
J. E. Kim and G. Carosi, arXiv:0807.3125 (2008).
%
\bibitem{pq} R. Peccei and H. Quinn, \prl{38}{1977}{1440} and
\prd{16}{1977}{1791}.
%
\bibitem{ww} S. Weinberg, \prl{40}{1978}{223};
F. Wilczek, \prl{40}{1978}{279}.
%
\bibitem{ksvz} J. E. Kim, \prl{43}{1979}{103};
M. A. Shifman, A. Vainstein and V. I. Zakharov, \npb{166}{1980}{493}.
%
\bibitem{dfsz} M. Dine, W. Fischler and M. Srednicki, \plb{104}{1981}{199};
A. P. Zhitnitskii, \sjp{31}{1980}{260}.
%
\bibitem{absik} L. F. Abbott and P. Sikivie, \plb{120}{1983}{133};
J. Preskill, M. Wise and F. Wilczek, \plb{120}{1983}{127};
M. Dine and W. Fischler, \plb{120}{1983}{137};
M. Turner, \prd{33}{1986}{889}; K. J. Bae, J. H. Huh and J. E. Kim, 
JCAP{\bf 0809} (2008) 005; L. Visinelli and P. Gondolo, \prd{80}{2009}{035024}.
%
%
\bibitem{nillesraby} H. P. Nilles and S. Raby, \npb{198}{1982}{102}.
%
\bibitem{steffen_rev} For recent reviews of axion/axino dark matter, see
F. Steffen, \epjc{59}{2009}{557};
L. Covi and J. E. Kim, \njp{11}{2009}{105003}.
%
\bibitem{axmass} T. Goto and M. Yamaguchi, \plb{276}{1992}{103}; 
E. J. Chun, J. E. Kim and H. P. Nilles, 
\plb{287}{1992}{123}.
%
\bibitem{rtw} K. Rajagopal, M. Turner and F. Wilczek, 
\npb{358}{1991}{447}.
%
\bibitem{cl} E. J. Chun and A. Lukas, \plb{357}{1995}{43}.
%
\bibitem{ckkr} L. Covi, J. E. Kim and L. Roszkowski, \prl{82}{1999}{4180}; 
L. Covi, H. B. Kim, J. E. Kim and L. Roszkowski, \jhep{0105}{2001}{033}.
L. Covi, L. Roszkowski and Small, \jhep{0207}{2002}{023}. 
%
\bibitem{fstw} A. Freitas, F. Steffen, N. Tajuddin and D. Wyler,
\plb{679}{2009}{270}.
%
\bibitem{cmssm} L. Covi, L. Roszkowski, R. Ruiz de Austri and M. Small,
\jhep{0406}{2004}{003};
A. Brandenburg, L. Covi, K. Hamaguchi, L. Roszkowski and
F. Steffen, \plb{617}{2005}{99}; K. Y. Choi, L. Roszkowski and
R. Ruiz de Austri, \jhep{0804}{2008}{016}.
%
\bibitem{axdm} H. Baer, A. Box and H. Summy, \jhep{0908}{2009}{080}.
%
\bibitem{msugra} A.~Chamseddine, R.~Arnowitt and P.~Nath, \prl{49}{1982}{970};
R.~Barbieri, S.~Ferrara and C.~Savoy, \plb {119}{1982}{343};
N.~Ohta, Prog. Theor. Phys. {\bf 70} (1983) 542; L. Hall,
J. Lykken and S. Weinberg, \prd {27}{1983}{2359}.
%
\bibitem{pierre} P. Sikivie and Q. Yang, \prl{103}{2009}{111301}.
%
\bibitem{mix} H. Baer, S. Kraml, S. Sekmen and H. Summy, \jhep{0803}{2008}{056\
};
H. Baer and H. Summy, \plb{666}{2008}{5};
H. Baer, M. Haider, S. Kraml,  S. Sekmen and H. Summy,
JCAP{\bf 0902} (2009) 002.
%
\bibitem{eo} J. Ellis and K. Olive, \plb{514}{2001}{114}.
%
\bibitem{isajet} F. Paige, S. Protopopescu, H. Baer and X. Tata, \hepph{0312045}; 
http://www.nhn.ou.edu/$\sim$isajet/
%
\bibitem{haber} H. E. Haber, R. Hempfling and A. Hoang, \zpc{75}{1996}{539}.
%
\bibitem{pbmz} D.Pierce, J. Bagger, K. Matchev and R. Zhang,
\npb{491}{1997}{3}.
%
\bibitem{isared} H. Baer, C. Balazs, A. Belyaev, \jhep{0203}{2002}{042}.
%
\bibitem{ltanb} H. Baer, C. H. Chen, M. Drees, F. Paige and X. Tata,
\prl{79}{1997}{986}.
%
\bibitem{hb_fp} K.~L.~Chan, U.~Chattopadhyay and P.~Nath, \prd{58}{1998}{096004\
};
J.~Feng, K.~Matchev and T.~Moroi, \prl{84}{2000}{2322} and
\prd{61}{2000}{075005}; see also
H.~Baer, C.~H.~Chen, F.~Paige and X.~Tata, \prd{52}{1995}{2746} and
\prd{53}{1996}{6241};
H.~Baer, C.~H.~Chen, M.~Drees, F.~Paige and X.~Tata, \prd{59}{1999}{055014};
for a model-independent approach, see
H.~Baer, T.~Krupovnickas, S.~Profumo and P.~Ullio, \jhep{0510}{2005}{020}.
%
\bibitem{stau} J.~Ellis, T.~Falk and K.~Olive, \plb{444}{1998}{367};
J.~Ellis, T.~Falk, K.~Olive and M.~Srednicki, \app{13}{2000}{181};
M.E.~G\'{o}mez, G.~Lazarides and C.~Pallis, \prd{61}{2000}{123512}
and \plb{487}{2000}{313};
A.~Lahanas, D.~V.~Nanopoulos and V.~Spanos, \prd{62}{2000}{023515};
R.~Arnowitt, B.~Dutta and Y.~Santoso, \npb{606}{2001}{59};
see also Ref.~\cite{isared}.
%
\bibitem{hfunnel} R.~Arnowitt and P.~Nath, \prl{70}{1993}{3696};
H.~Baer and M.~Brhlik, \prd{53}{1996}{597};
A.~Djouadi, M.~Drees and J.~Kneur, \plb{624}{2005}{60}.
%
\bibitem{bulk} H.~Goldberg, \prl {50}{1983}{1419};
J.~Ellis {\it et al.} \npb {238}{1984}{453}; P.~Nath and R.~Arnowitt,
\prl {70}{1993}{3696}; H.~Baer and M.~Brhlik, 
Ref.~\cite{hfunnel};
V.~Barger and C.~Kao, \prd{57}{1998}{3131}.
%
\bibitem{Afunnel} M.~Drees and M.~Nojiri, \prd{47}{1993}{376};
H.~Baer and M.~Brhlik, \prd{57}{1998}{567};
H.~Baer, M.~Brhlik, M.~Diaz, J.~Ferrandis, P.~Mercadante,
P.~Quintana and X.~Tata, \prd{63}{2001}{015007};
J.~Ellis, T.~Falk, G.~Ganis, K.~Olive and M.~Srednicki, \plb{510}{2001}{236};
L.~Roszkowski, R.~Ruiz de Austri and T.~Nihei, \jhep{0108}{2001}{024};
A.~Djouadi, M.~Drees and J.~L.~Kneur, \jhep{0108}{2001}{055};
A.~Lahanas and V.~Spanos, \epjc{23}{2002}{185}.
%
\bibitem{kimnilles} J. E. Kim and H. P. Nilles, \plb{138}{1984}{150}.
%
\bibitem{kohri} K. Kohri, T. Moroi and A. Yotsuyanagi, \prd{73}{2006}{123511};
for an update, see
M. Kawasaki, K. Kohri, T. Moroi and A. Yotsuyanagi, \prd{78}{2008}{065011};
see also J.~Pradler and F.~D.~Steffen, \plb{648}{2007}{224}.
%
\bibitem{sugmasses} S. Soni and H. A. Weldon, \plb{126}{1983}{215};
V.Kaplunovsky and J. Louis, \plb{306}{1993}{269};
A. Brignole, L. Ibanez and C. Munoz, \npb{422}{1994}{125}.
%
\bibitem{cnqw} For a recent analysis, see 
M. Carena, G. Nardini, M. Quiros and C. Wagner, \npb{812}{2009}{243}.
%
\bibitem{tev+stop} A.~A.~Affolder {\it et al.}  [CDF Collaboration],
  Phys.\ Rev.\ Lett.\  {\bf 84} (2000) 5273;
%
\bibitem{leptog} M. Fukugita and T. Yanagida, \plb{174}{1986}{45};
M. Luty, \prd{45}{1992}{455};
W. Buchm\"uller and M. Plumacher, \plb{389}{1996}{73} and \ijmpa{15}{2000}{5047};
R. Barbieri, P. Creminelli, A. Strumia and N. Tetradis, \npb{575}{2000}{61};
G. F. Giudice, A. Notari, M. Raidal, A. Riotto and A. Strumia, 
\npb{685}{2004}{89};
for a recent review, see W. Buchm\"uller, R. Peccei and T. Yanagida, 
\arnps{55}{2005}{311}.
%
\bibitem{buchm} W. Buchmuller, P. Di Bari and M. Plumacher, 
Annal. Phys. {\bf 315} (2005) 305.
%
\bibitem{NTlepto} G. Lazarides and Q. Shafi, \plb{258}{1991}{305};
K. Kumekawa, T. Moroi and T. Yanagida, \ptp{92}{1994}{437};
T. Asaka, K. Hamaguchi, M. Kawasaki and T. Yanagida, \plb{464}{1999}{12}.
%
%
%
\bibitem{imy} M. Ibe, T. Moroi and T. Yanagida, \plb{620}{2005}{9}.
%
\bibitem{ad} I.~Affleck and M.~Dine,
  Nucl.\ Phys.\  B {\bf 249} (1985) 361.
%
\bibitem{my} H. Murayama and T. Yanagida, \plb{322}{1994}{349};
M. Dine, L. Randall and S. Thomas, \npb{458}{1996}{291}.
%
\bibitem{jlm} K. Jedamzik, M. LeMoine and G. Moultaka,
JCAP{\bf 0607} (2006) 010.
%
\bibitem{steffen} A. Brandenburg and F.~Steffen,
JCAP{\bf 0408} (2004) 008.
%
\bibitem{kolbturner} R. Kolb and M. Turner, {\it The Early Universe} (Addison-Wesley, 1990).
%
\bibitem{njp} For a recent review of direct, indirect and collider detection
of neutralino dark matter, see H. Baer, E. K. Park and X. Tata, \njp{11}{2009}{105024}.
%
\bibitem{admx} 
L. Duffy {\it et al.}, \prl{95}{2005}{091304} and \prd{74}{2006}{012006};
for a review, see S. Asztalos, L. Rosenberg, K. van Bibber, P. Sikivie
and K. Zioutas, \arnps{56}{2006}{293}.
%
\bibitem{ewft} For some discussion of electroweak fine-tuning in supersymmetric models, 
see {\it e.g.} J. Ellis, K. Enqvist, D. V. Nanopoulos and F. Zwirner, 
\mpla{1}{1986}{57}; R. Barbieri and G. F. Giudice, \npb{306}{1988}{63};
G. Anderson and D. Castano, \plb{347}{1995}{300} and \prd{52}{1995}{1693};
J. L. Feng, K. Matchev and T. Moroi, \prl{84}{2000}{2322} and \prd{61}{2000}{075005}.  
%
\bibitem{wtn} N.~Arkani-Hamed, A.~Delgado and G.~F.~Giudice,
  \npb{741}{2006}{108}; H.~Baer, A.~Mustafayev, E.~K.~Park and X.~Tata,
JCAP {\bf 0701} (2007) 017 and \jhep{0805}{2008}{058}.

%
\end{thebibliography}
\end{document}